%% file: 15ACHAlinf.tex
\documentclass[preprint,12pt]{elsarticle}

\usepackage[T1]{fontenc}
\usepackage{graphicx}

\usepackage{amssymb}
\usepackage{amsmath}
\usepackage{paralist}
\usepackage{subfigure}
\usepackage{setspace}

\usepackage{algorithm}
\usepackage{algpseudocode}

\biboptions{sort&compress}

\input{vmr-symbols-vecbold}
\input{standard-macros}

\input{defs}

\newcommand{\primal}{$(\text{P}^{\,\varepsilon}_\infty)$}

\usepackage{color}

\newproof{IEEEproof}{Proof}

\journal{a Journal}

\begin{document}

\begin{frontmatter}

\title{{\bf Democratic Representations}}

\author{Christoph~Studer$^1$, Tom Goldstein$^2$, Wotao Yin$^3$, and Richard G.~Baraniuk$^4$} 


\address{$^1$School of Electrical and Computer Engineering, Cornell University, Ithaca, NY \\[0.05cm]
$^2$Department of Computer Science, University of Maryland, College Park, MD  \\[0.05cm]
$^3$Department of Mathematics, University of California, Los Angeles, CA \\[0.05cm]
$^4$Dept.~of Electrical and Computer Engineering, Rice University, Houston, TX}


\begin{abstract}
Minimization of the $\ell_{\infty}$ (or maximum) norm subject to a constraint that imposes consistency to an underdetermined system of linear equations finds use in a large number of practical applications, including vector quantization, approximate nearest neighbor search, peak-to-average power ratio (or ``crest factor'') reduction in communication systems, and peak force minimization in robotics and control.
This paper analyzes the fundamental properties of signal representations obtained by solving such a convex optimization problem. We develop bounds on the maximum magnitude of such representations using the uncertainty principle~(UP) introduced by Lyubarskii and Vershynin, and study the efficacy of $\ell_{\infty}$-norm-based dynamic range reduction.
Our analysis shows that matrices satisfying the UP, such as randomly subsampled Fourier or i.i.d.~Gaussian matrices, enable the computation of what we call \emph{democratic representations}, whose entries all have small and similar  magnitude, as well as low dynamic range.
To compute democratic representations at low computational complexity, we present two new, efficient convex optimization algorithms. 
We finally demonstrate the efficacy of democratic representations for dynamic range reduction in a DVB-T2-based broadcast system. 
\end{abstract}



\end{frontmatter}

\input{1introduction}
\input{2representations}

\input{3mainresults}

\input{4frames}

\input{5algorithm}

\input{6results}

\section{Conclusions}
\label{sec:conclusions}

In this paper, we have analyzed a host of fundamental properties of signal representations with minimum $\ellinf$ (or maximum) norm. 
Specifically, we have developed  properties on the magnitudes of such representations, and we characterized their peak-to-average power (PAPR) properties, which is of practical interest for OFDM-based communication systems.
We have demonstrated the existence of matrices for which \emph{democratic representations} with small $\ellinf$-norm and small PAPR exist.
We have furthermore developed two new and computationally efficient algorithms for solving~$(\text{P}^{\,\varepsilon}_\infty)$. 
To support our analysis, we have conducted a set of numerical experiments, which highlight that (i) Parseval frames lead to democratic representations with smaller $\ellinf$-norm compared to general frames and (ii) democratic representations offer tremendous PAPR reduction gains over existing approaches.

There are many avenues for follow-on research. An analytical characterization of the sharp phase transitions for~$(\text{P}^{\,\varepsilon}_\infty)$ observed in \fref{sec:simkashinbound}, e.g., using techniques developed in \cite{DT10,donoho2009}, is an interesting open research problem. 
In addition, the development of algorithms particularly suited for PAPR reduction in OFDM-based communication systems is left for future work. 
We also believe that assessing the efficacy of democratic representations in other practical applications, such as vector quantization, approximate nearest neighbor search, filter design, or robotics and control, is an interesting research direction.


%

  
\appendix

\input{appendix}

\section*{Acknowledgments}

Thanks to C.~Hegde, E.~G.~Larsson, A.~Maleki, K.~Mitra, G.~Pope, and A.~Sankaranarayanan for discussions on $\ell_\infty$-norm minimization. Thanks to Mr.\ Lan for verifying our proofs. 

The work of C.~Studer was supported in part by the Grants SNSF~PA00P2-134155 and NSF ECCS-1408006. 
The work of W.~Yin was supported by NSF Grant ECCS-1028790, ONR Grant N00014-08-1-1101, and ARO MURI W911NF-09-1-0383.
The work of C.~Studer, T.~Goldstein and R.~G.~Baraniuk was supported in part by the Grants NSF CCF-0431150, CCF-0728867, CCF-0926127, DARPA/ONR N66001-08-1-2065, N66001-11-1-4090, N66001-11-C-4092, ONR N00014-08-1-1112, N00014-10-1-0989, AFOSR FA9550-09-1-0432, ARO MURIs W911NF-07-1-0185 and W911NF-09-1-0383.


\bibliographystyle{elsarticle-num}

\bibliography{IEEEabrv,confs-jrnls,publishers,studer}

\end{document}

%% file: vmr-symbols-vecbold.tex
\usepackage{amssymb}
\usepackage{amsfonts}
\usepackage{mathrsfs}
\usepackage{xspace}
\usepackage{bm}
\usepackage{upgreek}

\newcommand{\safemath}[2]{\newcommand{#1}{\ensuremath{#2}\xspace}}

\safemath{\bma}{\mathbf{a}}
\safemath{\bmb}{\mathbf{b}}
\safemath{\bmc}{\mathbf{c}}
\safemath{\bmd}{\mathbf{d}}
\safemath{\bme}{\mathbf{e}}
\safemath{\bmf}{\mathbf{f}}
\safemath{\bmg}{\mathbf{g}}
\safemath{\bmh}{\mathbf{h}}
\safemath{\bmi}{\mathbf{i}}
\safemath{\bmj}{\mathbf{j}}
\safemath{\bmk}{\mathbf{k}}
\safemath{\bml}{\mathbf{l}}
\safemath{\bmm}{\mathbf{m}}
\safemath{\bmn}{\mathbf{n}}
\safemath{\bmo}{\mathbf{o}}
\safemath{\bmp}{\mathbf{p}}
\safemath{\bmq}{\mathbf{q}}
\safemath{\bmr}{\mathbf{r}}
\safemath{\bms}{\mathbf{s}}
\safemath{\bmt}{\mathbf{t}}
\safemath{\bmu}{\mathbf{u}}
\safemath{\bmv}{\mathbf{v}}
\safemath{\bmw}{\mathbf{w}}
\safemath{\bmx}{\mathbf{x}}
\safemath{\bmy}{\mathbf{y}}
\safemath{\bmz}{\mathbf{z}}
\safemath{\bmzero}{\mathbf{0}}
\safemath{\bmone}{\mathbf{1}}

\bmdefine{\biad}{a}
\bmdefine{\bibd}{b}
\bmdefine{\bicd}{c}
\bmdefine{\bidd}{d}
\bmdefine{\bied}{e}
\bmdefine{\bifd}{f}
\bmdefine{\bigd}{g}
\bmdefine{\bihd}{h}
\bmdefine{\biid}{i}
\bmdefine{\bijd}{j}
\bmdefine{\bikd}{k}
\bmdefine{\bild}{l}
\bmdefine{\bimd}{m}
\bmdefine{\bind}{n}
\bmdefine{\biod}{o}
\bmdefine{\bipd}{p}
\bmdefine{\biqd}{q}
\bmdefine{\bird}{r}
\bmdefine{\bisd}{s}
\bmdefine{\bitd}{t}
\bmdefine{\biud}{u}
\bmdefine{\bivd}{v}
\bmdefine{\biwd}{w}
\bmdefine{\bixd}{x}
\bmdefine{\biyd}{y}
\bmdefine{\bizd}{z}

\bmdefine{\bixid}{\xi}
\bmdefine{\bilambdad}{\lambda}
\bmdefine{\bimud}{\mu}
\bmdefine{\bithetad}{\theta}
\bmdefine{\biphid}{\phi}
\bmdefine{\bideltad}{\delta}

\safemath{\bmia}{\biad}
\safemath{\bmib}{\bibd}
\safemath{\bmic}{\bicd}
\safemath{\bmid}{\bidd}
\safemath{\bmie}{\bied}
\safemath{\bmif}{\bifd}
\safemath{\bmig}{\bigd}
\safemath{\bmih}{\bihd}
\safemath{\bmii}{\biid}
\safemath{\bmij}{\bijd}
\safemath{\bmik}{\bikd}
\safemath{\bmil}{\bild}
\safemath{\bmim}{\bimd}
\safemath{\bmin}{\bind}
\safemath{\bmio}{\biod}
\safemath{\bmip}{\bipd}
\safemath{\bmiq}{\biqd}
\safemath{\bmir}{\bird}
\safemath{\bmis}{\bisd}
\safemath{\bmit}{\bitd}
\safemath{\bmiu}{\biud}
\safemath{\bmiv}{\bivd}
\safemath{\bmiw}{\biwd}
\safemath{\bmix}{\bixd}
\safemath{\bmiy}{\biyd}
\safemath{\bmiz}{\bizd}

\safemath{\bmxi}{\bixid}
\safemath{\bmlambda}{\bilambdad}
\safemath{\bmmu}{\bimud}
\safemath{\bmtheta}{\bithetad}
\safemath{\bmphi}{\biphid}
\safemath{\bmdelta}{\bideltad}

\safemath{\bA}{\mathbf{A}}
\safemath{\bB}{\mathbf{B}}
\safemath{\bC}{\mathbf{C}}
\safemath{\bD}{\mathbf{D}}
\safemath{\bE}{\mathbf{E}}
\safemath{\bF}{\mathbf{F}}
\safemath{\bG}{\mathbf{G}}
\safemath{\bH}{\mathbf{H}}
\safemath{\bI}{\mathbf{I}}
\safemath{\bJ}{\mathbf{J}}
\safemath{\bK}{\mathbf{K}}
\safemath{\bL}{\mathbf{L}}
\safemath{\bM}{\mathbf{M}}
\safemath{\bN}{\mathbf{N}}
\safemath{\bO}{\mathbf{O}}
\safemath{\bP}{\mathbf{P}}
\safemath{\bQ}{\mathbf{Q}}
\safemath{\bR}{\mathbf{R}}
\safemath{\bS}{\mathbf{S}}
\safemath{\bT}{\mathbf{T}}
\safemath{\bU}{\mathbf{U}}
\safemath{\bV}{\mathbf{V}}
\safemath{\bW}{\mathbf{W}}
\safemath{\bX}{\mathbf{X}}
\safemath{\bY}{\mathbf{Y}}
\safemath{\bZ}{\mathbf{Z}}

\safemath{\bZero}{\mathbf{0}}
\safemath{\bOne}{\mathbf{1}}
\safemath{\bDelta}{\mathbf{\Delta}}
\safemath{\bLambda}{\mathbf{\UpLambda}}
\safemath{\bPhi}{\mathbf{\Upphi}}
\safemath{\bSigma}{\mathbf{\Upsigma}}
\safemath{\bOmega}{\mathbf{\Upomega}}
\safemath{\bTheta}{\mathbf{\Uptheta}}

\bmdefine{\biAd}{A}
\bmdefine{\biBd}{B}
\bmdefine{\biCd}{C}
\bmdefine{\biDd}{D}
\bmdefine{\biEd}{E}
\bmdefine{\biFd}{F}
\bmdefine{\biGd}{G}
\bmdefine{\biHd}{H}
\bmdefine{\biId}{I}
\bmdefine{\biJd}{J}
\bmdefine{\biKd}{K}
\bmdefine{\biLd}{L}
\bmdefine{\biMd}{M}
\bmdefine{\biOd}{N}
\bmdefine{\biPd}{O}
\bmdefine{\biQd}{P}
\bmdefine{\biRd}{R}
\bmdefine{\biSd}{S}
\bmdefine{\biTd}{T}
\bmdefine{\biUd}{U}
\bmdefine{\biVd}{V}
\bmdefine{\biWd}{W}
\bmdefine{\biXd}{X}
\bmdefine{\biYd}{Y}
\bmdefine{\biZd}{Z}

\bmdefine{\biDelta}{\Delta}
\bmdefine{\biLambda}{\Lambda}
\bmdefine{\biPhi}{\Phi}
\bmdefine{\biSigma}{\Sigma}
\bmdefine{\biOmega}{\Omega}
\bmdefine{\biTheta}{\Theta}

\safemath{\bimA}{\biAd}
\safemath{\bimB}{\biBd}
\safemath{\bimC}{\biCd}
\safemath{\bimD}{\biDd}
\safemath{\bimE}{\biEd}
\safemath{\bimF}{\biFd}
\safemath{\bimG}{\biGd}
\safemath{\bimH}{\biHd}
\safemath{\bimI}{\biId}
\safemath{\bimJ}{\biJd}
\safemath{\bimK}{\biKd}
\safemath{\bimL}{\biLd}
\safemath{\bimM}{\biMd}
\safemath{\bimN}{\biNd}
\safemath{\bimO}{\biOd}
\safemath{\bimP}{\biPd}
\safemath{\bimQ}{\biQd}
\safemath{\bimR}{\biRd}
\safemath{\bimS}{\biSd}
\safemath{\bimT}{\biTd}
\safemath{\bimU}{\biUd}
\safemath{\bimV}{\biVd}
\safemath{\bimW}{\biWd}
\safemath{\bimX}{\biXd}
\safemath{\bimY}{\biYd}
\safemath{\bimZ}{\biZd}

\safemath{\bimDelta}{\biDelta}
\safemath{\bimLambda}{\biLambda}
\safemath{\bimPhi}{\biPhi}
\safemath{\bimSigma}{\biSigma}
\safemath{\bimOmega}{\biOmega}
\safemath{\bimTheta}{\biTheta}

\safemath{\setA}{\mathcal{A}}
\safemath{\setB}{\mathcal{B}}
\safemath{\setC}{\mathcal{C}}
\safemath{\setD}{\mathcal{D}}
\safemath{\setE}{\mathcal{E}}
\safemath{\setF}{\mathcal{F}}
\safemath{\setG}{\mathcal{G}}
\safemath{\setH}{\mathcal{H}}
\safemath{\setI}{\mathcal{I}}
\safemath{\setJ}{\mathcal{J}}
\safemath{\setK}{\mathcal{K}}
\safemath{\setL}{\mathcal{L}}
\safemath{\setM}{\mathcal{M}}
\safemath{\setN}{\mathcal{N}}
\safemath{\setO}{\mathcal{O}}
\safemath{\setP}{\mathcal{P}}
\safemath{\setQ}{\mathcal{Q}}
\safemath{\setR}{\mathcal{R}}
\safemath{\setS}{\mathcal{S}}
\safemath{\setT}{\mathcal{T}}
\safemath{\setU}{\mathcal{U}}
\safemath{\setV}{\mathcal{V}}
\safemath{\setW}{\mathcal{W}}
\safemath{\setX}{\mathcal{X}}
\safemath{\setY}{\mathcal{Y}}
\safemath{\setZ}{\mathcal{Z}}
\safemath{\emptySet}{\varnothing}

\safemath{\colA}{\mathscr{A}}
\safemath{\colB}{\mathscr{B}}
\safemath{\colC}{\mathscr{C}}
\safemath{\colD}{\mathscr{D}}
\safemath{\colE}{\mathscr{E}}
\safemath{\colF}{\mathscr{F}}
\safemath{\colG}{\mathscr{G}}
\safemath{\colH}{\mathscr{H}}
\safemath{\colI}{\mathscr{I}}
\safemath{\colJ}{\mathscr{J}}
\safemath{\colK}{\mathscr{K}}
\safemath{\colL}{\mathscr{L}}
\safemath{\colM}{\mathscr{M}}
\safemath{\colN}{\mathscr{N}}
\safemath{\colO}{\mathscr{O}}
\safemath{\colP}{\mathscr{P}}
\safemath{\colQ}{\mathscr{Q}}
\safemath{\colR}{\mathscr{R}}
\safemath{\colS}{\mathscr{S}}
\safemath{\colT}{\mathscr{T}}
\safemath{\colU}{\mathscr{U}}
\safemath{\colV}{\mathscr{V}}
\safemath{\colW}{\mathscr{W}}
\safemath{\colX}{\mathscr{X}}
\safemath{\colY}{\mathscr{Y}}
\safemath{\colZ}{\mathscr{Z}}

\safemath{\opA}{\mathbb{A}}
\safemath{\opB}{\mathbb{B}}
\safemath{\opC}{\mathbb{C}}
\safemath{\opD}{\mathbb{D}}
\safemath{\opE}{\mathbb{E}}
\safemath{\opF}{\mathbb{F}}
\safemath{\opG}{\mathbb{G}}
\safemath{\opH}{\mathbb{H}}
\safemath{\opI}{\mathbb{I}}
\safemath{\opJ}{\mathbb{J}}
\safemath{\opK}{\mathbb{K}}
\safemath{\opL}{\mathbb{L}}
\safemath{\opM}{\mathbb{M}}
\safemath{\opN}{\mathbb{N}}
\safemath{\opO}{\mathbb{O}}
\safemath{\opP}{\mathbb{P}}
\safemath{\opQ}{\mathbb{Q}}
\safemath{\opR}{\mathbb{R}}
\safemath{\opS}{\mathbb{S}}
\safemath{\opT}{\mathbb{T}}
\safemath{\opU}{\mathbb{U}}
\safemath{\opV}{\mathbb{V}}
\safemath{\opW}{\mathbb{W}}
\safemath{\opX}{\mathbb{X}}
\safemath{\opY}{\mathbb{Y}}
\safemath{\opZ}{\mathbb{Z}}
\safemath{\opZero}{\mathbb{O}}
\safemath{\identityop}{\opI}

\safemath{\veca}{\bma}
\safemath{\vecb}{\bmb}
\safemath{\vecc}{\bmc}
\safemath{\vecd}{\bmd}
\safemath{\vece}{\bme}
\safemath{\vecf}{\bmf}
\safemath{\vecg}{\bmg}
\safemath{\vech}{\bmh}
\safemath{\veci}{\bmi}
\safemath{\vecj}{\bmj}
\safemath{\veck}{\bmk}
\safemath{\vecl}{\bml}
\safemath{\vecm}{\bmm}
\safemath{\vecn}{\bmn}
\safemath{\veco}{\bmo}
\safemath{\vecp}{\bmp}
\safemath{\vecq}{\bmq}
\safemath{\vecr}{\bmr}
\safemath{\vecs}{\bms}
\safemath{\vect}{\bmt}
\safemath{\vecu}{\bmu}
\safemath{\vecv}{\bmv}
\safemath{\vecw}{\bmw}
\safemath{\vecx}{\bmx}
\safemath{\vecy}{\bmy}
\safemath{\vecz}{\bmz}

\safemath{\veczero}{\bmzero}
\safemath{\vecone}{\bmone}
\safemath{\vecxi}{\bmxi}
\safemath{\veclambda}{\bmlambda}
\safemath{\vecmu}{\bmmu}
\safemath{\vectheta}{\bmtheta}
\safemath{\vecphi}{\bmphi}
\safemath{\vecdelta}{\bmdelta}

\safemath{\matA}{\bA}
\safemath{\matB}{\bB}
\safemath{\matC}{\bC}
\safemath{\matD}{\bD}
\safemath{\matE}{\bE}
\safemath{\matF}{\bF}
\safemath{\matG}{\bG}
\safemath{\matH}{\bH}
\safemath{\matI}{\bI}
\safemath{\matJ}{\bJ}
\safemath{\matK}{\bK}
\safemath{\matL}{\bL}
\safemath{\matM}{\bM}
\safemath{\matN}{\bN}
\safemath{\matO}{\bO}
\safemath{\matP}{\bP}
\safemath{\matQ}{\bQ}
\safemath{\matR}{\bR}
\safemath{\matS}{\bS}
\safemath{\matT}{\bT}
\safemath{\matU}{\bU}
\safemath{\matV}{\bV}
\safemath{\matW}{\bW}
\safemath{\matX}{\bX}
\safemath{\matY}{\bY}
\safemath{\matZ}{\bZ}
\safemath{\matzero}{\bmzero}

\safemath{\matDelta}{\bDelta}
\safemath{\matLambda}{\bLambda}
\safemath{\matPhi}{\bPhi}
\safemath{\matSigma}{\bSigma}
\safemath{\matOmega}{\bOmega}
\safemath{\matTheta}{\bTheta}

\safemath{\matidentity}{\matI}
\safemath{\matone}{\matO}

\safemath{\rnda}{A}
\safemath{\rndb}{B}
\safemath{\rndc}{C}
\safemath{\rndd}{D}
\safemath{\rnde}{E}
\safemath{\rndf}{F}
\safemath{\rndg}{G}
\safemath{\rndh}{H}
\safemath{\rndi}{I}
\safemath{\rndj}{J}
\safemath{\rndk}{K}
\safemath{\rndl}{L}
\safemath{\rndm}{M}
\safemath{\rndn}{N}
\safemath{\rndo}{O}
\safemath{\rndp}{P}
\safemath{\rndq}{Q}
\safemath{\rndr}{R}
\safemath{\rnds}{S}
\safemath{\rndt}{T}
\safemath{\rndu}{U}
\safemath{\rndv}{V}
\safemath{\rndw}{W}
\safemath{\rndx}{X}
\safemath{\rndy}{Y}
\safemath{\rndz}{Z}

\safemath{\rveca}{\bimA}
\safemath{\rvecb}{\bimB}
\safemath{\rvecc}{\bimC}
\safemath{\rvecd}{\bimD}
\safemath{\rvece}{\bimE}
\safemath{\rvecf}{\bimF}
\safemath{\rvecg}{\bimG}
\safemath{\rvech}{\bimH}
\safemath{\rveci}{\bimI}
\safemath{\rvecj}{\bimJ}
\safemath{\rveck}{\bimK}
\safemath{\rvecl}{\bimL}
\safemath{\rvecm}{\bimM}
\safemath{\rvecn}{\bimN}
\safemath{\rveco}{\bomO}
\safemath{\rvecp}{\bimP}
\safemath{\rvecq}{\bimQ}
\safemath{\rvecr}{\bimR}
\safemath{\rvecs}{\bimS}
\safemath{\rvect}{\bimT}
\safemath{\rvecu}{\bimU}
\safemath{\rvecv}{\bimV}
\safemath{\rvecw}{\bimW}
\safemath{\rvecx}{\bimX}
\safemath{\rvecy}{\bimY}
\safemath{\rvecz}{\bimZ}

\safemath{\rvecxi}{\bmxi}
\safemath{\rveclambda}{\bmlambda}
\safemath{\rvecmu}{\bmmu}
\safemath{\rvectheta}{\bmtheta}
\safemath{\rvecphi}{\bmphi}

\safemath{\rmatA}{\bimA}
\safemath{\rmatB}{\bimB}
\safemath{\rmatC}{\bimC}
\safemath{\rmatD}{\bimD}
\safemath{\rmatE}{\bimE}
\safemath{\rmatF}{\bimF}
\safemath{\rmatG}{\bimG}
\safemath{\rmatH}{\bimH}
\safemath{\rmatI}{\bimI}
\safemath{\rmatJ}{\bimJ}
\safemath{\rmatK}{\bimK}
\safemath{\rmatL}{\bimL}
\safemath{\rmatM}{\bimM}
\safemath{\rmatN}{\bimN}
\safemath{\rmatO}{\bimO}
\safemath{\rmatP}{\bimP}
\safemath{\rmatQ}{\bimQ}
\safemath{\rmatR}{\bimR}
\safemath{\rmatS}{\bimS}
\safemath{\rmatT}{\bimT}
\safemath{\rmatU}{\bimU}
\safemath{\rmatV}{\bimV}
\safemath{\rmatW}{\bimW}
\safemath{\rmatX}{\bimX}
\safemath{\rmatY}{\bimY}
\safemath{\rmatZ}{\bimZ}

\safemath{\rmatDelta}{\bimDelta}
\safemath{\rmatLambda}{\bimLambda}
\safemath{\rmatPhi}{\bimPhi}
\safemath{\rmatSigma}{\bimSigma}
\safemath{\rmatOmega}{\bimOmega}
\safemath{\rmatTheta}{\bimTheta}

%% file: standard-macros.tex
\usepackage{amssymb}
\usepackage{amsfonts}
\usepackage{mathrsfs}
\usepackage{xspace}
\usepackage{bm}
\usepackage{fancyref}
\usepackage{textcomp}

\usepackage{multirow}
\usepackage{stmaryrd}

\newenvironment{textbmatrix}{	\setlength{\arraycolsep}{2.5pt}%
								\big[\begin{matrix}}{\end{matrix}\big]%
								\raisebox{0.08ex}{\vphantom{M}}}

\def\be{\begin{equation}}
\def\ee{\end{equation}}
\def\een{\nonumber \end{equation}}
\def\mat{\begin{bmatrix}}
\def\emat{\end{bmatrix}}
\def\btm{\begin{textbmatrix}}
\def\etm{\end{textbmatrix}}

\def\ba#1\ea{\begin{align}#1\end{align}}
\def\bas#1\eas{\begin{align*}#1\end{align*}}
\def\bs#1\es{\begin{split}#1\end{split}} 
\def\bg#1\eg{\begin{gather}#1\end{gather}}
\def\bml#1\eml{\begin{multline}#1\end{multline}}
\def\bi#1\ei{\begin{itemize}#1\end{itemize}}

\newcommand{\lefto}{\mathopen{}\left}

\DeclareMathOperator{\rank}{rank}			
\DeclareMathOperator{\sign}{sign}			

\newcommand{\inftytilde}{{\widetilde\infty}}

\newcommand{\abs}[1]{\lefto\lvert#1\right\rvert}		

\newcommand{\vecnorm}[1]{\lefto\lVert#1\right\rVert}		
\newcommand{\herm}[1]{\ensuremath{#1^{H}}} 	

\safemath{\dirac}{\delta}					
\safemath{\krond}{\dirac}					

\safemath{\upto}{\uparrow}
\safemath{\downto}{\downarrow}
\safemath{\iu}{j}							
\safemath{\ev}{\lambda}						
\safemath{\hilseqspace}{l^{2}}				
\newcommand{\banachfunspace}[1]{\setL^{#1}}	
\safemath{\hilfunspace}{\banachfunspace{2}}	

\newcommand{\ceil}[1]{\lceil #1 \rceil}

\safemath{\SNR}{\textsf{SNR}} 				
\safemath{\PAR}{\textsf{PAR}} 				
\safemath{\No}{N_0}							
\safemath{\Es}{E_s}							
\safemath{\Eb}{E_b}							
\safemath{\EbNo}{\frac{\Eb}{\No}}
\safemath{\EsNo}{\frac{\Es}{\No}}

\DeclareMathOperator{\CHop}{\ensuremath{\opH}} 
\safemath{\tvir}{\rndh_{\CHop}}				
\safemath{\tvtf}{\rndl_{\CHop}}				
\safemath{\spf}{\rnds_{\CHop}}				
\safemath{\bff}{H_{\CHop}}					

\safemath{\ircf}{r_{h}}						
\safemath{\tftvcf}{r_{s}}					
\safemath{\tfcf}{r_{l}}						
\safemath{\bfcf}{r_{H}}						

\safemath{\tcorr}{c_h}						
\safemath{\scf}{c_{s}}						
\safemath{\tfcorr}{c_{l}}					
\safemath{\fcorr}{c_{H}}						

\safemath{\mi}{I}							
\safemath{\capacity}{C}						

\safemath{\normal}{\mathcal{N}}			
\safemath{\jpg}{\mathcal{CN}}			
\safemath{\mchain}{\leftrightarrow}		

\safemath{\dB}{\,\mathrm{dB}}
\safemath{\dBm}{\,\mathrm{dBm}}
\safemath{\Hz}{\,\mathrm{Hz}}
\safemath{\kHz}{\,\mathrm{kHz}}
\safemath{\MHz}{\,\mathrm{MHz}}
\safemath{\GHz}{\,\mathrm{GHz}}
\safemath{\s}{\,\mathrm{s}}
\safemath{\ms}{\,\mathrm{ms}}
\safemath{\mus}{\,\mathrm{\text{\textmu}s}}
\safemath{\ns}{\,\mathrm{ns}}
\safemath{\ps}{\,\mathrm{ps}}
\safemath{\meter}{\,\mathrm{m}}
\safemath{\mm}{\,\mathrm{mm}}
\safemath{\cm}{\,\mathrm{cm}}
\safemath{\m}{\,\mathrm{m}}
\safemath{\W}{\,\mathrm{W}}
\safemath{\mW}{\, \mathrm{mW}}
\safemath{\J}{\,\mathrm{J}}
\safemath{\K}{\,\mathrm{K}}
\safemath{\bit}{\,\mathrm{bit}}
\safemath{\nat}{\,\mathrm{nat}}

\safemath{\define}{\triangleq}			

\safemath{\equivalent}{\sim}
\safemath{\distas}{\sim}					
\safemath{\sdiff}{\Delta}				

\safemath{\reals}{\mathbb{R}}
\safemath{\positivereals}{\reals_{+}}
\safemath{\integers}{\mathbb{Z}}
\safemath{\posint}{\integers_{+}}
\safemath{\naturals}{\mathbb{N}}
\safemath{\posnaturals}{\naturals_{+}}
\safemath{\complexset}{\mathbb{C}}
\safemath{\rationals}{\mathbb{Q}}

\newcommand*{\fancyrefapplabelprefix}{app}		
\newcommand*{\fancyrefthmlabelprefix}{thm}		
\newcommand*{\fancyreflemlabelprefix}{lem}		
\newcommand*{\fancyrefcorlabelprefix}{cor}		
\newcommand*{\fancyrefdeflabelprefix}{def}		
\newcommand*{\fancyrefproplabelprefix}{prop}		
\newcommand*{\fancyrefobslabelprefix}{obs}		
\newcommand*{\fancyrefexmpllabelprefix}{exmpl}
\newcommand*{\fancyrefalglabelprefix}{alg}		

\frefformat{vario}{\fancyrefseclabelprefix}{Section~#1}
\frefformat{vario}{\fancyrefthmlabelprefix}{Theorem~#1}
\frefformat{vario}{\fancyreflemlabelprefix}{Lemma~#1}
\frefformat{vario}{\fancyrefcorlabelprefix}{Corollary~#1}
\frefformat{vario}{\fancyrefdeflabelprefix}{Definition~#1}
\frefformat{vario}{\fancyrefobslabelprefix}{Observation~#1}
\frefformat{vario}{\fancyreffiglabelprefix}{Fig.~#1}
\frefformat{vario}{\fancyrefapplabelprefix}{#1} 
\frefformat{vario}{\fancyrefeqlabelprefix}{(#1)}
\frefformat{vario}{\fancyrefproplabelprefix}{Proposition~#1}
\frefformat{vario}{\fancyrefexmpllabelprefix}{Example~#1}
\frefformat{vario}{\fancyrefalglabelprefix}{Algorithm~#1}

%% file: defs.tex
 \newtheorem{thm}{Theorem}
 
 \newtheorem{defi}{Definition}
 \newtheorem{lem}[thm]{Lemma} 

\safemath{\dictab}{[\,\dicta\,\,\dictb\,]}

\safemath{\ysig}{\bmy}
\safemath{\ysighat}{\hat{\ysig}}
\safemath{\ysigdim}{M}
\safemath{\xsig}{\bmx}
\safemath{\xsigdim}{N}
\safemath{\nx}{n_x}
\safemath{\zsig}{\bmz}
\safemath{\zsigdim}{\ysigdim}
\safemath{\rsig}{\bmr}
\safemath{\Adict}{\bA}
\safemath{\Adicttilde}{\widetilde{\Adict}}
\safemath{\Adictdim}{\outputdim\times\xsigdim}
\safemath{\avec}{\bma}
\safemath{\avectilde}{\tilde{\avec}}
\safemath{\Bdict}{\bB}
\safemath{\Bdicttilde}{\widetilde{\Bdict}}
\safemath{\Cdict}{\bC}
\safemath{\cvec}{\bmc}
\safemath{\Ddict}{\bD}
\safemath{\Ddictdim}{\ysigdim\times\xsigdim}
\safemath{\dvec}{\bmd}
\safemath{\Ddicttilde}{\widetilde{\bD}}
\safemath{\Bonb}{\bB}
\safemath{\bvec}{\bmb}
\safemath{\Bonbdim}{\ysigdim\times\ysigdim}
\safemath{\noise}{\bmn}
\safemath{\noisedim}{\ysigim}
\safemath{\err}{\bme}
\safemath{\errdim}{\ysigdim}
\safemath{\errset}{\setE}
\safemath{\nerr}{n_e}
\safemath{\delop}{\bP_\errset}
\safemath{\delopc}{\bP_{{\errset}^c}}

\safemath{\cplxi}{\imath}
\safemath{\cplxj}{\jmath}

\safemath{\dict}{\matD}
\safemath{\inputdim}{N}		
\safemath{\outputdim}{M}		
\safemath{\sparsity}{S}	
\safemath{\inputdimA}{{N_a}}	
\safemath{\inputdimB}{{N_b}}	
\safemath{\elemA}{{n_a}}	
\safemath{\elemB}{{n_b}}	
\safemath{\resA}{\matR_a}	
\safemath{\resB}{\matR_b}	
\safemath{\subD}{\matS} 
\safemath{\subA}{\matS_a} 
\safemath{\subB}{\matS_b} 
\safemath{\dicta}{\matA} 	
\safemath{\dictb}{\matB} 	
\safemath{\hollowS}{H}
\safemath{\hollowA}{H_a}
\safemath{\hollowB}{H_b}
\safemath{\cross}{Z}
\safemath{\coh}{\mu_d}			
\safemath{\coha}{\mu_a}			
\safemath{\cohb}{\mu_b}			
\safemath{\mubs}{\nu}	
\safemath{\cohm}{\mu_m} 
\safemath{\dictset}{\setD}	
\safemath{\dictsetp}{\dictset(\coh,\coha,\cohb)}	
\safemath{\dictsetgen}{\dictset_\text{gen}}
\safemath{\dictsetgenp}{\dictsetgen(\coh)}
\safemath{\dictsetonb}{\dictset_\text{onb}}
\safemath{\dictsetonbp}{\dictsetonb(\coh)}

\safemath{\leftside}{U}
\safemath{\rightsideA}{R_a}
\safemath{\rightsideB}{R_b}

\safemath{\indexS}{\setI_S} 

\safemath{\na}{n_a}			
\safemath{\nb}{n_b}			
\safemath{\coeffa}{p_i}	
\safemath{\coeffb}{q_j}	
\safemath{\seta}{\setP}		
\safemath{\setb}{\setQ}     
\safemath{\setw}{\setW}	
\safemath{\setz}{\setZ}	
\safemath{\cola}{\veca}		
\safemath{\colb}{\vecb}		
\safemath{\cold}{\vecd}		
\safemath{\inputvec}{\vecx} 	
\safemath{\error}{\vece}	
\safemath{\noiseout}{\vecz} 	
\safemath{\inputvecel}{x}
\safemath{\inputveca}{\vecx_a}
\safemath{\inputvecb}{\vecx_b}
\safemath{\outputvec}{\vecy}	
\safemath{\lambdamin}{\lambda_{\mathrm{min}}}

\newcommand{\normtwo}[1]{\vecnorm{#1}_2}
\newcommand{\normone}[1]{\vecnorm{#1}_1}

\newcommand{\norminf}[1]{\vecnorm{#1}_\infty}
\newcommand{\norminftilde}[1]{\vecnorm{#1}_{\widetilde\infty}}
\newcommand{\spectralnorm}[1]{\vecnorm{#1}_{2,2}}
\safemath{\elltwo}{\ell_2}
\safemath{\ellone}{\ell_1}
\safemath{\ellzero}{\ell_0}
\safemath{\ellinf}{\ell_\infty}
\safemath{\ellinftilde}{\ell_{\widetilde\infty}}
\safemath{\licard}{Z(\coh,\coha,\cohb)}
\safemath{\xsol}{\hat{x}}
\safemath{\xbord}{x_b}		
\safemath{\xstat}{x_s}		
\safemath{\xstatLone}{\tilde{x}_s}
\safemath{\order}{\mathcal{O}} 
\safemath{\scales}{\Theta} 
\safemath{\ones}{\mathbf{1}} 
\safemath{\zeroes}{\mathbf{0}} 
\safemath{\thlone}{\kappa(\coh,\cohb)} 
\safemath{\constoneA}{\delta} 
\safemath{\constoneB}{\epsilon} 
\safemath{\nlarge}{L}				   
\safemath{\sumlarge}{S_\nlarge}
\safemath{\maxlarger}{P_\nlarge}	   
\safemath{\Pzero}{\textrm{P0}}	
\safemath{\Pone}{\textrm{P1}}
\safemath{\vecfir}{\vecw}			 
\safemath{\vecsec}{\vecz}
\safemath{\elvecfir}{w}              
\safemath{\elvecsec}{z}				 
\safemath{\nlargefir}{n}
\safemath{\normout}{\gamma}
\safemath{\auxfun}{h}
\safemath{\supp}{\textrm{supp}}

\safemath{\indexa}{\ell}
\safemath{\indexb}{r}
\safemath{\indexc}{i}
\safemath{\indexd}{j}

\safemath{\project}{P}

%% file: 1introduction.tex
\section{Introduction}
\label{sec:intro}

%
In this paper, we analyze the properties of the solutions~$\dot\inputvec\in\complexset^N$ to the following convex minimization problem:
\begin{align*}
(\text{P}^{\,\varepsilon}_\infty) \quad \underset{\tilde\vecx\in\complexset^N}{\text{minimize}}\,\, \norminf{\tilde\vecx} \quad \text{subject to}\,\, \normtwo{\vecy-\dict\tilde\vecx} \leq \varepsilon.
\end{align*}
Here, the vector $\vecy\in\complexset^M$ denotes the \emph{signal} to be represented, \mbox{$\dict\in\complexset^{M\times N}$} is an overcomplete matrix (often called frame or dictionary) with \mbox{$M< N$}, and the real-valued approximation parameter $\varepsilon\geq0$ determines the accuracy of the \emph{signal representation}~$\dot\vecx$.

As demonstrated in~\cite{LV10}, certain matrices~\dict enable the computation of  signal representations~$\dot\vecx$ whose entries all have magnitudes of the order~$1/\sqrt{N}$. 
Since for such representations each entry is of approximately the same importance, we call them  \emph{democratic}.\footnote{Other names for {democratic representations} have been proposed in the literature. The paper~\cite{LV10} uses ``{Kashin representations},'' whereas~\cite{Fuchs11} uses both, ``{spread representations}'' and ``{anti-sparse representations}.'' We also note that \cite{CD02} used the term ``democracy'' for quantized representations where the individual bits have ``equal-weight'' in the context of sigma-delta conversion. Here, the signal representations $\dot\vecx$ are, in general, neither binary-valued nor quantized.}

\begin{figure}[tp]
\centering
\subfigure[]{\includegraphics[width=0.7\columnwidth]{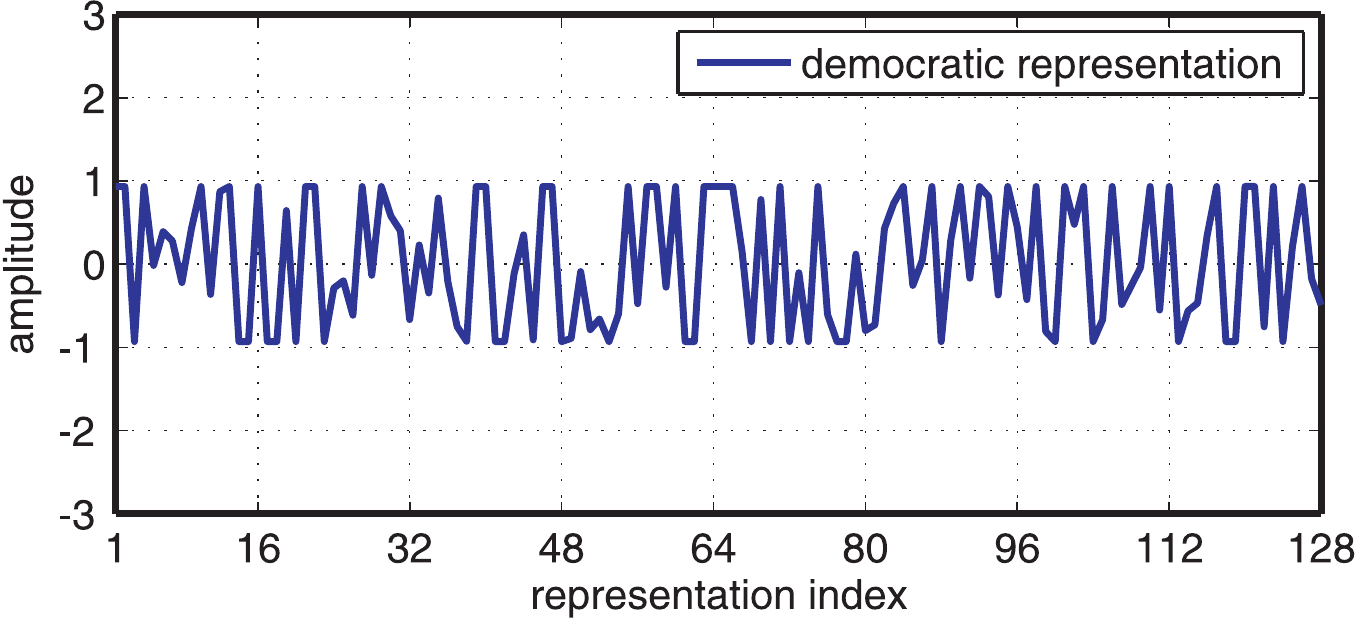}}
\subfigure[]{\includegraphics[width=0.7\columnwidth]{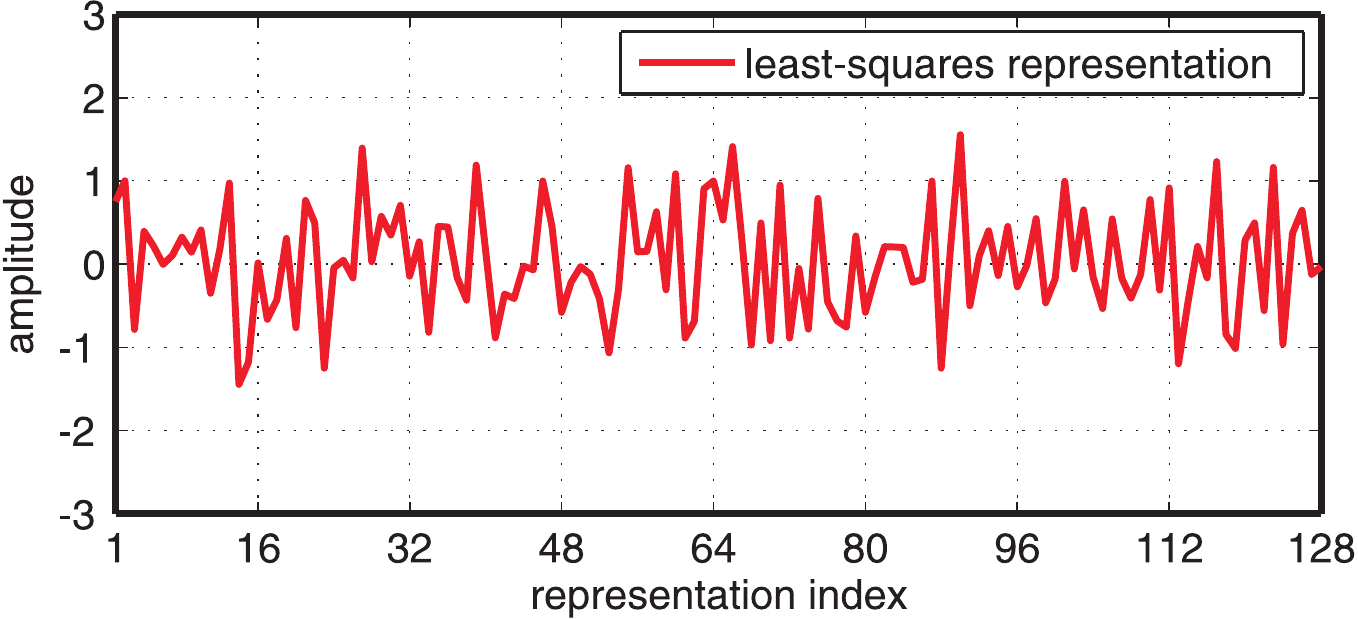}}
\subfigure[]{\includegraphics[width=0.7\columnwidth]{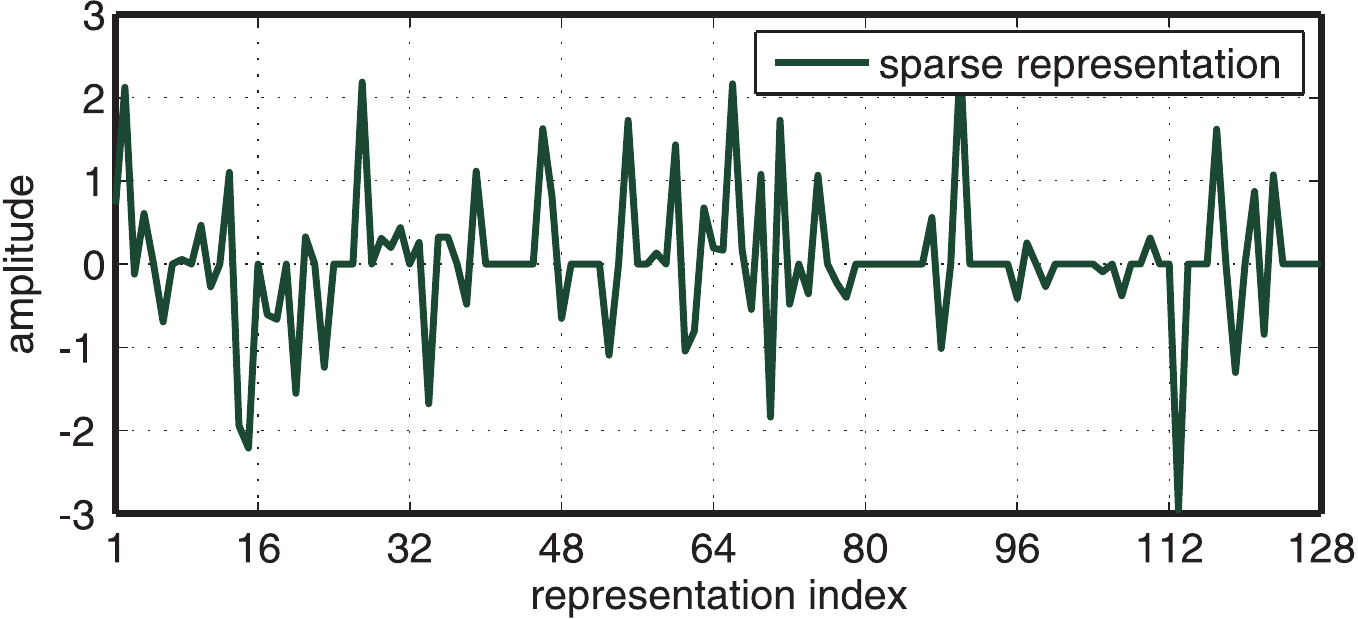}}
\caption{There are, in general, infinitely many ways to represent a signal as a linear combination of columns from an overcomplete matrix. In this example, we compare the representations of a given signal vector that are optimal according to three different criteria, i.e.,  minimum $\ellinf$, $\ell_2$, and $\ell_1$-norm, using an overcomplete $64\times128$ subsampled DCT matrix; (a) democratic ($\ellinf$-norm) representation ($11.1$\,dB PAPR); (b) least-squares ($\ell_2$-norm) representation ($16.2$\,dB PAPR); (c) sparse ($\ellone$-norm) representation ($29.9$\,dB PAPR). We see that the democratic representation has (i) most of its entries' magnitudes at the same maximum level and (ii) very small dynamic range (also in terms of the PAPR).}
\label{fig:example}
\end{figure}

Figure~\ref{fig:example} shows an example of three different representations of the same signal $\vecy$ using the columns of a subsampled discrete cosine transform (DCT) matrix.\footnote{The entries of the vector $\vecy$ are generated from a  zero-mean  i.i.d.\ Gaussian distribution with unit variance; the row indices of the DCT matrix have been chosen uniformly at random. All representations are computed via problems of the form~$(\text{P}^{\,\varepsilon}_\infty)$ with the $\ellinf$, $\ell_2$, and $\ellone$-norm for the democratic, least-squares, and sparse representation, respectively, and we set $\varepsilon=0$.}
In contrast to the (popular) least-squares and sparse representation, most of the entries of the democratic representation obtained via $(\text{P}^{\,\varepsilon}_\infty)$ have the same (low) maximum magnitude.
As a consequence of this particular magnitude property, the problem~$(\text{P}^{\,\varepsilon}_\infty)$ and the resulting signal representations feature prominently in a variety of practical applications.

\subsection{Application Examples}
\label{sec:appexamples}

%
%
\subsubsection{Vector quantization}
Element-wise quantization of  democratic representations affects all entries of~$\dot\vecx$ equally, which renders them less susceptible to quantization noise  compared with  direct quantization of the signal vector~$\vecy$~\cite{LV10}. 
Moreover, the corruption of a few entries of $\dot\vecx$ results in only a small error and, therefore, computing $\vecy=\dict\dot\vecx$ after, e.g., transmission over an unreliable communication channel~\cite{PK05} or storage in unreliable memory cells~\cite{Novak2010}, enables one to obtain a robust estimate of the signal vector~$\vecy$.

\subsubsection{Peak-to-average power ratio (PAPR) reduction} 
Wireless communication systems employing orthogonal frequency division multiplexing (OFDM) typically require linear and power-inefficient radio-frequency (RF) components (e.g., power amplifiers) to avoid unwanted signal distortions or out-of-band radiation, as OFDM signals are prone to exhibit a large peak-to-average power ratio (PAPR) (also called the ``crest factor'')~\cite{NP00}; see \fref{eq:pardefinition} for the PAPR definition used in this paper.
%
%
By allocating certain unused OFDM tones, which is known as tone reservation~\cite{IS09}, or by exploiting the excess degrees-of-freedom in large-scale multi-antenna wireless systems (often called massive MIMO systems), one can transmit {democratic representations}, which significantly reduce the PAPR~\cite{SL12}. 
Hence, transmitting  democratic representations, instead of conventional (unprocessed) OFDM signals, substantially alleviates the need for expensive and power-inefficient RF components.
The example in \fref{fig:example} confirms that the democratic representation exhibits substantially lower PAPR compared to a least-squares or sparse representation.

%

%
%
%
%
%
%
%
\subsubsection{Approximate nearest neighbor search} Signal representations obtained from the problem $(\text{P}^{\,\varepsilon}_\infty)$ also find use in the identification of approximate nearest neighbors in high-dimensional spaces~\cite{JFF11}.
The underlying idea is to compute a representation~$\dot\vecx$ for the query vector $\vecy$. For certain matrices~$\bD$, the resulting representations are democratic and hence, resemble to an antipodal signal for which most coefficients take the values $-\alpha$ or $+\alpha$ for some \mbox{$\alpha>0$}; see the democratic representation example in \fref{fig:example} where $\alpha\approx1$. This property of the coefficients of $\dot\vecx$ can then be used to efficiently find approximate nearest vectors in an $N$-dimensional Hamming space.
\subsubsection{Robotics and control} Kinematically redundant robots or manipulators admit infinitely many inverse solutions. Certain applications require a solution that minimizes the \emph{maximum} force,  acceleration, torque, or joint velocity, rather than minimizing the energy or power.
Hence, in many practical situations, one is typically interested in solving problems of the form~$(\text{P}^{\,\varepsilon}_\infty)$ to obtain democratic representations with provably small $\ellinf$-norm, rather than minimum-energy (or least-squares) representations (the democratic representation in \fref{fig:example} has significantly smaller $
\ellinf$-norm compared to the other two representations); corresponding practical application examples have been discussed in~\cite{Cadzow71,DW97,ZWX02}.
%

%

\subsubsection{Recovery conditions for sparse signal recovery}
As shown in~\cite{ZYY13}, the $\ell_\infty$-norm of the representation obtained by solving a specific instance of $(\text{P}^{\,\varepsilon}_\infty)$ with $\varepsilon=0$ can be used to verify uniqueness and robustness conditions for $\ell_1$-norm-based (analysis and synthesis) sparse signal recovery problems. Such recovery conditions are of particular interest in the emerging fields of sparse signal recovery~\cite{tropp2004,eladbook2010,SB11} and compressive sensing~(CS)~\cite{donoho2006,candes2006c,candes2008a,baraniuk2008a}. 

\subsection{What About Signal Recovery?}

The problem $(\text{P}^{\,\varepsilon}_\infty)$ with $\varepsilon=0$ can be used to \emph{recover} antipodal (or binary-valued) signals, i.e., vectors with coefficients belonging to the set $\{-\alpha,+\alpha\}$ for $\alpha>0$ from the underdetermined system of linear equations $\vecy=\dict\vecx$, provided that the matrix $\dict$ meets certain conditions~\cite{DT10,MR11,chandrasekaran2012convex,oymak2013simple}.
The main focus here is, however, on (i) properties of {democratic representations} having minimal $\ellinf$-norm and small dynamic range, and (ii) their efficient computation, rather than on the {recovery} of a given antipodal vector~$\vecx$ from the set of linear equations  $\vecy=\dict\vecx$ and corresponding uniqueness conditions. 
We refer the interested reader to  \cite{DT10,MR11,chandrasekaran2012convex,oymak2013simple} for the details on antipodal signal recovery via $\ell_\infty$-norm minimization in noiseless and noisy settings.


\fussy

\sloppy
\subsection{Relevant Prior Art on $\ellinf$-Norm Minimization}

Results for minimizing the maximum amplitude of continuos, real-valued signals subject to linear constraints reach back to the 1960s, when Neustadt~\cite{Neustadt66} studied the so-called  \emph{minimum-effort control problem}.
In 1971, Cadzow proposed a corresponding practicable algorithm suitable for low-dimensional systems, where he proposed to solve the following real-valued, convex \mbox{$\ellinf$-norm} minimization problem~\cite{Cadzow71}:
\begin{align*}
(\text{P}_\infty) \quad \underset{\tilde\vecx\in\reals^N}{\text{minimize}}\,\, \norminf{\tilde\vecx} \quad \text{subject to}\,\, \vecy=\dict\tilde\vecx.
\end{align*}
Note that this problem coincides to a real-valued version of $(\text{P}^{\,\varepsilon}_\infty)$ with $\varepsilon=0$.
Specifically,  in \cite{Cadzow71} it was shown that, for a large class of matrices \dict, there exists a solution $\dot\vecx$ to $(\text{P}_\infty)$ for which a dominant portion of the entries' magnitudes correspond to $\norminf{\dot\vecx}$, whereas only a small fraction of the entries may have smaller magnitude; this result has been rediscovered recently by Fuchs in~\cite{Fuchs11}.

Another line of research that characterizes signal representations~$\vecx$ with small (but not necessarily minimal) $\ellinf$-norm subject to $\vecy=\dict\vecx$ have been established in 2010 by Lyubarskii and Vershynin~\cite{LV10}.
In particular, \cite{LV10} proves the existence of matrices~$\dict$ with arbitrarily small redundancy parameter \mbox{$\lambda=N/M>1$} for which every signal vector $\vecy$ has a democratic representation~$\vecx$ satisfying 
\begin{align} \label{eq:kashinrepdef}
\norminf{\vecx} \leq \frac{K}{\sqrt{N}} \normtwo{\vecy}.
\end{align}
Here, $K$ is a (preferably small) constant that only depends on the redundancy parameter~$\lambda$.
The existence of such signal representations for certain sets of matrices can either be shown using fundamental results obtained by Kashin~\cite{Kashin77}, Garnaev and Gluskin~\cite{GG84}, or by analyzing the signal representations obtained via the iterative algorithm proposed in~\cite{LV10}.
The latter (constructive) approach relies on an uncertainty principle (UP) for the matrix \dict, which establishes a fundamental connection between the constant $K$ in~\fref{eq:kashinrepdef} and sensing matrices commonly used in  sparse signal recovery and~CS~\cite{donoho2006,candes2006d}.

\fussy

\subsection{Contributions}

In this paper, we derive and investigate a host of fundamental properties for signal representations~$\dot\vecx$ obtained from the $\ellinf$-norm minimization problem~$(\text{P}^{\,\varepsilon}_\infty)$.
In particular, we analyze its Lagrange dual problem to derive a refined and more general version of the bound on the $\ellinf$-norm of the signal representation~$\dot\vecx$ established in~\cite{LV10}.
We characterize magnitude properties of the signal representations~$\dot\vecx$ obtained by solving~$(\text{P}^{\,\varepsilon}_\infty)$, and we develop bounds on the resulting PAPR, which is of particular interest in OFDM-based communication systems.
As a byproduct of our analysis, we present the Lagrange duals to a variety of optimization problems, such as $\ellone$-norm minimization, which is often used for sparse signal recovery and compressive sensing.
We then discuss classes of matrices that enable the computation of \emph{democratic representations} via~$(\text{P}^{\,\varepsilon}_\infty)$.
Furthermore, we develop two computationally efficient algorithms to solve~$(\text{P}^{\,\varepsilon}_\infty)$, referred to as CRAM (short for \underline{c}onvex \underline{r}eduction of \underline{am}plitudes) and CRAMP (short for CRAM for Parseval frames). CRAM is suitable for arbitrary matrices $\bD$ and approximation parameters $\varepsilon\in[0,\infty)$, whereas CRAMP exhibits lower complexity than CRAM for $\varepsilon=0$ and Parseval frames. We provide numerical experiments to support our analysis and conclude by demonstrating the efficacy of democratic representations for PAPR reduction in a DVB-T2-based broadcast system~\cite{DVBT2}.

\sloppy

\subsection{Notation}
\label{sec:notation}
Lowercase boldface letters stand for column vectors and uppercase
boldface letters designate matrices. For a matrix~\bA, we denote its conjugate transpose and spectral norm
by $\herm{\bA}$ and $\spectralnorm{\bA}=\sqrt{\lambda_\text{max}(\bA^H\bA)}$, respectively, where $\lambda_\text{max}(\bA^H\bA)$ denotes the maximum eigenvalue of $\bA^H\bA$.
We use $\bZero_{M\times N}$ and $\bOne_{M\times N}$ to denote the all-zeros and all-ones matrix of dimension $M\times N$, respectively.
The $k^\text{th}$ entry of a vector $\bma$ is designated by $[\veca]_k$, and  $\Re\{\veca\}$ and $\Im\{\veca\}$ represent its real and imaginary part, respectively.
We define the $\ell_p$-norm of  the vector~$\veca\in\complexset^N$ as follows:
\begin{align*}
 \|{\veca}\|_p = \left\{\begin{array}{ll}
\left(\sum_{k=1}^N\abs{[\veca]_k}^p\right)^{\!1/p} & \text{if} \,\,\, 1\leq p < \infty \\[0.2cm]
\max_{k\in\{1,\ldots,N\}}\abs{[\veca]_k} & \text{if} \,\,\, p=\infty.
\end{array}\right.
\end{align*}
We also make use of the (non-standard) $\ellinftilde$-norm~\cite{Seethaler10} defined as
$\norminftilde{\bma}=\max\!\big\{\norminf{\Re\{\veca\}}\!,\norminf{\Im\{\veca\}}\!\big\}$. 
The notation $\dot\vecx$ is used to refer to the solutions to the problem~$(\text{P}^{\,\varepsilon}_\infty)$.
Sets are designated by uppercase Greek letters; the cardinality of the set $\Omega$ is $\abs{\Omega}$.
The notation $\supp(\veca)$ designates to the support set of the vector $\veca$, i.e., the set of indices associated to non-zero entries in $\veca$.
%
%
The sign (or phase) of a complex-valued scalar $x\in\complexset$ is defined as
\begin{align*}
 \sign(x) = \left\{\begin{array}{ll}
x/\abs{x} & \text{if} \,\,\, x\neq0 \\[0.15cm]
0 & \text{if} \,\,\, x=0.
\end{array}\right.
\end{align*}
We use $\sign(\veca)$ and $\mathrm{abs}(\veca)$ to denote the entry-wise application of the sign function and absolute value to the vector~$\veca$, respectively.

\fussy

\subsection{Organization of the Paper}

The remainder of the paper is organized as follows. Section~\ref{sec:frames} introduces the essentials of frames and the uncertainty property (UP). In \fref{sec:theory}, we develop the concept of democratic representations. Our main results are detailed in \fref{sec:properties}.
\fref{sec:UPsatisfyingmatrices} reviews suitable classes of matrices that satisfy the UP and enable the computation of democratic representations. \fref{sec:algorithmyipee} develops computationally efficient algorithms for solving~$(\text{P}^{\,\varepsilon}_\infty)$. \fref{sec:simulation} provides numerical experiments and showcases the efficacy of democratic representations for PAPR reduction. 
We conclude in \fref{sec:conclusions}. Most proofs are relegated to the Appendices.

%% file: 2representations.tex

\section{Frames and the Uncertainty Principle}  \label{sec:frames} 

\subsection{Frames}
We often require the over-complete matrix \mbox{$\dict\in\complexset^{M\times N}$} with \mbox{$M<N$} to satisfy the following property.

\begin{defi}[Frame~\cite{MB12}] \label{def:frame}
A matrix \mbox{$\dict\in\complexset^{M\times N}$} with \mbox{$M\leq N$} is called a \emph{frame} if
\begin{align*} 
A\normtwo{\vecw}^2 \leq \normtwo{\dict^H\vecw}^2 \leq B\normtwo{\vecw}^2
\end{align*} 
holds for any vector $\vecw\in\complexset^M$ with $A\in\reals$, $B\in\reals$, and \mbox{$0<A\leq B<\infty$}. 
\end{defi}

\sloppy 
The tightest possible constants $A$ and $B$ are called the lower and upper \emph{frame bounds}, respectively. 
The frame~$\dict$ is called a \emph{tight frame} if \mbox{$A=B$}. Furthermore, if \mbox{$A=B=1$}, then~$\dict$ is a \emph{Parseval frame}~\cite{MB12}. In what follows, we exclusively study the finite-dimensional setting (i.e., where $M,N<\infty$) and thus $B<\infty$. Further, because our frame definition requires~$A$ to be strictly positive,~$\bD$ is guaranteed to be full rank.   Thus, \primal{} is feasible for any frame $\bD$ and~$\varepsilon\geq0$. 

\subsection{Full-Spark Frames}
\label{sec:fullsparkframes}
The next definition is concerned with the \emph{spark} of a frame, which represents the cardinality of the smallest subset of linearly dependent frame columns.
\begin{defi}[Full-spark frame~\cite{ACM12}] \label{def:fullsparkframe}
A frame \mbox{$\dict\in\complexset^{M\times N}$} is called a \emph{full-spark frame} if the columns of 
every $M\times M$ sub-matrix of $\bD$ are linearly independent.
\end{defi} 

\fussy

Full-spark frames have spark \mbox{$M+1$} and are ubiquitous in sparse signal recovery and CS (see \cite{ACM12} for a review).
Even though verifying the full-spark property of an arbitrary matrix is, in general, a hard problem~\cite{Pfetsch2012}, many frames are known to be full spark.  For example, any subset of rows from a Fourier matrix of prime dimension forms a full-spark frame \cite{ACM12,PT96}. Further, Vandermonde matrices with  \mbox{$M\leq N$} having distinct basis entries are known to be full spark frames~\cite{Fuchs05,ACM12}. In addition, randomized constructions also exist that generate full-spark frames with high probability.  In particular, if the entries of an $M\times N$ matrix with $M<N$ are generated from independent continuous random variables, then the resulting matrix is a full-spark frame with probability one (see~\cite{Blumensath07} for a formal proof).

\subsection{The Uncertainty Principle (UP)} \label{sec:uncertaintyPrinciple}

Several of the results derived in this paper rely upon the uncertainty principle (UP) introduced in~\cite{LV10}.
\begin{defi}[Uncertainty principle~\cite{LV10}] \label{def:upforframes}
We say that the frame $\dict\in\complexset^{M\times N}$ satisfies the UP with parameters $\eta\in\reals^+$ and $\delta\in(0,1)$ if
\begin{align}
\normtwo{\dict\vecx} \leq \eta \normtwo{\vecx} \label{eq:up}
\end{align}
holds for all (sparse) vectors $\vecx\in\complexset^N$ satisfying $\abs{\supp(\vecx)}\leq\delta N$.   
\end{defi}

We emphasize that \eqref{eq:up} is trivially satisfied for $\eta=\spectralnorm{\bD}$ and for arbitrary vectors with $\abs{\supp(\vecx)}\leq N$.
However, as in~\cite{LV10}, we are particularly interested in frames satisfying the UP with parameters \mbox{$\eta < \spectralnorm{\bD}$} and $\delta<1$. For simplicity, we say that frames satisfying definition \eqref{def:upforframes}  with such non-trivial parameters ``satisfy the UP.''

Verifying the UP with parameters~$\eta<\spectralnorm{\bD}$ and $\delta<1$ for a given frame~\dict requires, in general, a combinatorial search over all $\delta N$-sparse vectors~\cite{Pfetsch2012}. 
Nevertheless, many classes of frames are known to satisfy the UP with high probability (see~\cite{LV10} and  \fref{sec:UPsatisfyingmatrices} for more details).
%
We finally note that frames satisfying the UP are strongly related to sensing matrices with small restricted isometry constants; such matrices play a central role in~CS~\cite{donoho2006,candes2006d,baraniuk2008a}. 
%

\section{Democratic Representations}
\label{sec:theory}

We next introduce the concept of \emph{democratic representations} and define the \emph{democracy constants}. 
\subsection{Democratic Representations}

For $M<N$ and $\varepsilon<\normtwo{\vecy}$, there exist, in general, an infinite number of representations $\vecx$  for a given signal vector $\vecy$ that satisfy \mbox{$\normtwo{\vecy-\dict\vecx}\leq\varepsilon$}.\footnote{Note in the case $\varepsilon\geq\normtwo{\vecy}$, the problem $(\text{P}^{\,\varepsilon}_\infty)$ returns the all-zeros vector and, hence, practically relevant choices of $\varepsilon$ are in the range $0\leq\varepsilon<\normtwo{\vecy}$.}
 In the remainder of the paper, we are particularly interested in representations for which every entry $x_i$, $i=1,\ldots,N$ is of approximately the same importance. In particular, we seek so-called \emph{democratic representations}, which have provably small $\ellinf$-norm and for which all magnitudes are approximately equal. 
%
%
%
In order to make the concept of democratic representations more formal, we use the following definition.

\begin{defi}[Democracy constants] \label{def:democracyconstants}
Let $\dict\in\complexset^{M\times N}$ be a given frame. 
Assume we obtain a signal representation~$\dot\vecx$ for every vector $\vecy\in\complexset^M$ by solving $(\text{P}^{\,\varepsilon}_\infty)$ with $\varepsilon\leq\normtwo{\vecy}$.
We define the {\em lower} and  {\em upper democracy constants} $K_\text{l}\in\reals$ and \mbox{$K_\text{u}\in\reals$} to be the largest and smallest constants for which  
\begin{align} \label{eq:kashinbounds}
\frac{K_\text{l}}{\sqrt{N}}\big(\normtwo{\vecy}-\varepsilon\big)
 \leq \norminf{\dot\vecx} \leq \frac{K_\text{u}}{\sqrt{N}}\big(\normtwo{\vecy}-\varepsilon\big)
\end{align}
holds for every pair $\dot\vecx$ and \mbox{$\vecy$}, and for any $0\leq\varepsilon\leq\normtwo{\vecy}$.
\end{defi}

We note that the democracy constants $K_\text{l}$ and $K_\text{u}$ depend {\em only} on properties of the frame~$\dict$  and the fact that all signal representations~$\dot\vecx$ are obtained via $(\text{P}^{\,\varepsilon}_\infty)$, and \emph{not} on the signal vector~$\vecy$.
Note that \fref{def:democracyconstants} enables us to analyze a generalized and refined setting of the special case \fref{eq:kashinrepdef} studied in~\cite{LV10} (see our results in \fref{sec:properties}).

In what follows, we are interested in (i) classes of frames for which  the lower and upper democracy constants  $K_\text{l}$, $K_\text{u}$ are both close to $1$, and (ii) computationally efficient algorithms that provably deliver such signal representations.
In particular, if~$K_\text{l}\approx1\approx K_\text{u}$, then all signal representations~$\dot\vecx$ have similar $\ellinf$-norm and every entry will have a maximum magnitude of $1/\sqrt{N}$ (assuming $\varepsilon=0$ and $\normtwo{\vecy}=1$). 
Since this property {\em evenly} spreads the signal vector's energy $\normtwo{\vecy}$ across all entries of~$\dot\vecx$,  we call such representations \emph{democratic}.

\begin{defi}[Democratic representations] \label{def:democraticrepresentations}
Let $\dict\in\complexset^{M\times N}$ be a given frame. 
If the associated democracy constants $K_\text{l}$ and \mbox{$K_\text{u}$} are both close to~$1$, then the signal representations $\dot\vecx$ obtained by $(\text{P}^{\,\varepsilon}_\infty)$ are called \emph{democratic representations}.
\end{defi}

%



%

\subsection{Computing Democratic Representations}

%
%

In order to compute representations $\vecx$ having small (but not necessarily minimal) $\ellinf$-norm subject to the set of linear equations $\vecy=\dict\vecx$, one can use the iterative algorithm proposed in~\cite{LV10}. This method efficiently computes such representations for real-valued and approximate Parseval frames, i.e., frames $\dict\in\reals^{M\times N}$ satisfying the UP in~\cite{LV10} with Frame bounds $A=1-\xi$ and $B=1+\xi$ for some small $\xi\geq0$. 
However, the algorithm in \cite{LV10}
\begin{inparaenum}[(i)]
\item does not solve~$(\text{P}^{\,\varepsilon}_\infty)$ and is, in general, not guaranteed to find representations~$\dot\vecx$ having the \emph{smallest} $\ellinf$-norm, 
\item requires knowledge of the UP parameters $\eta$, $\delta$, 
\item was introduced for real-valued systems,\footnote{A corresponding generalization of the algorithm in \cite{LV10} to the complex-valued case is straightforward.} and 
\item is only guaranteed to converge for approximate Parseval frames.
\end{inparaenum}
Moreover, if one is interested in \emph{approximate} representations~$\vecx$ for which $\normtwo{\vecy-\dict\vecx} >0$ rather than in \emph{perfect} representations satisfying $\vecy=\dict\vecx$, the algorithm in \cite{LV10} must be modified accordingly.
In order to overcome the limitations of the algorithm in~\cite{LV10}, we propose to directly solve the convex problem $(\text{P}^{\,\varepsilon}_\infty)$ instead; \fref{sec:algorithmyipee} will detail two corresponding (and computationally efficient) algorithms.
%
%

%
%
%

%% file: 3mainresults.tex

\section{Main Results}
\label{sec:properties}

We now analyze several key properties of signal representations $\dot\vecx$ obtained from solving $(\text{P}^{\,\varepsilon}_\infty)$.  \fref{sec:extremevalues} studies magnitude properties of the solutions to $(\text{P}^{\,\varepsilon}_\infty)$. \fref{sec:lagrangedual} introduces the Lagrange dual problem to~$(\text{P}^{\,\varepsilon}_\infty)$, which is key in the proofs of Sections \ref{sec:lowerdemocracy} and \ref{sec:upperkashinbound}, where we develop bounds on the lower and upper democracy constants $K_\text{l}$ and $K_\text{u}$, respectively. \fref{sec:PAR} analyzes the PAPR characteristics of democratic representations, and \fref{sec:devotedtodominikseethaler} outlines an extension of our results to the $\ell_{\widetilde{\infty}}$-norm.

\subsection{Extreme Values of Solutions to  $(\text{P}^{\,\varepsilon}_\infty)$  }
\label{sec:extremevalues}

The magnitudes of signal representations obtained via $(\text{P}^{\,\varepsilon}_\infty)$ exhibit specific and practically relevant  properties. 
To study them, we need the following definition. 
\begin{defi}[Extreme and moderate entries]
Given a vector $\vecx\in \complexset^N$, we call an entry $x_i$  \emph{extreme} if  \mbox{$|x_i| = \|\vecx\|_\infty$}; we further call an entry $x_i$ \emph{moderate} if \mbox{$|x_i| < \|\vecx\|_\infty$}. 
\end{defi}
  
Without any specific assumptions on the overcomplete matrix~$\bD$ (apart from being full-rank), we next show that there always \emph{exists} a solution to~\primal{} with a large portion of  extreme entries (see \fref{app:demsExist} for the proof). In words, a democratic representation is one with a large portion of extreme entries. 

\begin{lem}[Democratic representations exist] \label{lem:demsExist}
For any  full-rank matrix $\bD\in\complexset^{M\times N}$ with $M\leq N$, the problem \primal{} admits a solution $\dot\vecx$ with at least $N-M+1$ extreme entries. 
\end{lem}

This result implies that there exist signal representations~$\dot\vecx$ for which a large number of (extreme) entries have equal magnitude. In particular, by increasing the redundancy $\lambda=N/M$ of $\bD$, the problem~$(\text{P}^{\,\varepsilon}_\infty)$ admits representations for which the number of extreme values is arbitrarily close to~$N$.

The next result shows that---given the matrix~$\bD$ is a full-spark frame---\emph{every} solution to the problem~\primal{} has a bounded minimal number of extreme entries (see \fref{app:alwaysDem} for the proof).
\begin{lem}[All representations are democratic] \label{lem:alwaysDem}
If the frame $\bD$ has full spark, then every solution to~\primal{} has at least $N-M+1$ extreme entries.
\end{lem}

This result implies that for full-spark frames with large redundancy $\lambda=M/N$, every solution to \primal{} is a democratic representation (with most entries being extreme). 
As noted in \fref{sec:fullsparkframes}, a large number of deterministic and random constructions of full-spark frames are known. Hence, solving \primal{} allows the computation of democratic representations for a large number of frames. 
In addition, \fref{lem:alwaysDem} enables us to obtain the following $\ell_\infty$/$\ell_2$-norm inequality.

\begin{thm}[Democratic $\ell_\infty$/$\ell_2$-norm inequality] \label{thm:norminequ}
If $\bD$ is a full-spark frame, then 
\begin{align} \label{eq:fancypantsnormbound}
\sqrt{N-M+1} \norminf{\dot\vecx} \leq \normtwo{\dot\vecx}
\end{align}
holds for every representation $\dot\bmx$ obtained by solving \primal{}.
\end{thm}
\begin{IEEEproof}
The proof immediately follows from \fref{lem:alwaysDem} and the straightforward inequality $\normtwo{\dot\vecx}^2=\sum_{i=1}^{N}|\dot x_i|^2\geq(N-M+1)\norminf{\dot\vecx}^2$.
\end{IEEEproof}

We note that the norm inequality \eqref{eq:fancypantsnormbound} is stronger than the standard norm bound $\norminf{\vecx}\leq\normtwo{\vecx}$ for $M<N$ (which holds for arbitrary vectors $\vecx$). More importantly, as we show below in \fref{sec:PAR}, the refined norm inequality  \eqref{eq:fancypantsnormbound} is particularly useful for characterizing the limits of PAPR reduction methods that rely on democratic representations.

We conclude by noting that  results related to Lemmata \ref{lem:demsExist} and \ref{lem:alwaysDem} for the special problem~\primal{} with $\varepsilon=0$ have been developed in the literature~\cite{Fuchs11,Cadzow71}. In particular,~\cite{Cadzow71} establishes bounds on the minimum number of entries that satisfy \mbox{$x_i\leq\norminf{\vecx}$} (i.e., entries that are not necessarily extremal). This result, however, does not allow us to extract bounds on the number of extremal values, which is in contrast to Lemmata~\ref{lem:demsExist} and \ref{lem:alwaysDem}.
Reference~\cite{Fuchs11} mentions that signal representations obtained from $(\text{P}_\infty)$ have, in general, exactly \mbox{$N-M+1$} extreme entries. This result, however, is stated without proof and, more importantly, without explicitly specifying conditions on the classes of the matrices for which it is supposed to hold.

\subsection{Lagrange Dual Problem}
\label{sec:lagrangedual}

In order to derive bounds on the lower and upper democracy constants $K_\text{l}$ and~$K_\text{u}$, respectively, and to study the PAPR behavior of solutions to $(\text{P}^{\,\varepsilon}_\infty)$, we make use of the following theorem (see \fref{app:primaldualproblems} for the proof).
\begin{thm}[Lagrange dual problem] \label{thm:primaldualproblems}
Let the  $\ell_p$-norm  primal problem (with $1\leq p \leq \infty$) be
\begin{align*}
(\text{P}^{\,\varepsilon}_p) \quad \underset{\tilde\vecx\in\complexset^N}{\text{minimize}}\,\, \|{\tilde\vecx}\|_p \quad \text{subject to}\,\, \normtwo{\vecy-\dict\tilde\vecx} \leq \varepsilon.
\end{align*}
Then, the corresponding Lagrange dual problem is given by 
\begin{align*}
(\text{D}^{\,\varepsilon}_p) \quad 
\left\{\begin{array}{ll}
\underset{\tilde\vecz\in\complexset^M}{\text{maximize}}& \Re\!\left(\vecy^H\tilde\vecz\right) - \varepsilon \normtwo{\tilde\vecz}\\[0.2cm]
\text{subject to} & \|\dict^H\tilde\vecz\|_d\leq1
\end{array}\right.
\end{align*}
with $1/p+1/d=1$; for $p=1$ we have $q=\infty$ and vice versa. The norm $\|\cdot\|_d$ corresponds to the dual norm of~$\|\cdot\|_p$. 
\end{thm}

Note that \fref{thm:primaldualproblems} includes not only the Lagrange dual to the problem $(\text{P}^{\,\varepsilon}_\infty)$, but also other frequently studied optimization problems, such as the Lagrangian dual to~$(\text{P}^{\,\varepsilon}_1)$, which is often used for sparse signal recovery or~CS.

\subsection{Bound on Lower Democracy Constant $K_\text{l}$}
\label{sec:lowerdemocracy}

In order to characterize the lower democracy constant $K_l$ for a given frame~\bD, we next derive a corresponding lower bound.

\begin{lem}[Lower democracy bound] \label{lem:lowerkashinbound}
Let $\dict\in\complexset^{M\times N}$ be a frame with upper frame bound $B$. Then, every vector $\vecy\in\complexset^M$ admits a signal representation $\dot\vecx$ with the following lower bound $\widetilde{K}_\text{l} $ on the lower democracy constant~$K_l$ (see \fref{app:lowerkashinbound} for the proof): 
\begin{align} \label{eq:lowerkashinbound}
\widetilde{K}_\text{l} = \frac{1}{\sqrt{B}} \leq K_\text{l}.
\end{align}
\end{lem}
%

\sloppy 

The representations $\dot\vecx$ obtained from the problem~$(\text{P}^{\,\varepsilon}_\infty)$ are guaranteed to satisfy \fref{eq:lowerkashinbound}, as the proof of \fref{lem:lowerkashinbound} exploits properties of its solution.
In the special case $\varepsilon=0$ and for Parseval frames, \fref{lem:lowerkashinbound} guarantees that all vectors~$\vecy$ admit a signal representation satisfying 
\begin{align*} 
\frac{\normtwo{\vecy}}{\sqrt{N}} \leq \norminf{\dot\vecx}.
\end{align*}
This lower bound was established previously in \cite[Obs.~2.1b]{LV10}. \fref{lem:lowerkashinbound} generalizes this result to arbitrary frames and to representations for which $\varepsilon>0$. 
%
%
It is furthermore interesting to observe that the lower democracy bound in~\fref{eq:lowerkashinbound} only depends on the upper frame constant $B$ (and implicitly on the fact that frames satisfy $A>0$); this is in contrast to the bound on the upper democracy constant $K_\text{u}$ derived next.

\fussy

\subsection{Bound on Upper Democracy Constant $K_\text{u}$}
\label{sec:upperkashinbound}

In order to characterize the upper democracy constant $K_u$, we next derive an upper bound $\widetilde{K}_\text{u}$ on $K_\text{u}$ by using the uncertainty principle (UP) for frames (see \fref{app:proof_mainresult} for the proof).
\begin{thm}[Upper democracy bound] \label{thm:main_result}
Let $\dict\in\complexset^{M\times N}$ be a frame with frame bounds $A$, $B$ that satisfies the uncertainty principle (UP) with parameters $\eta$, $\delta$. Then, every signal vector~$\vecy$ admits a signal representation $\dot\vecx$ with the following upper bound on the upper democracy constant:
\begin{align} \label{eq:maincondition}
{K}_\text{u} \leq \widetilde{K}_\text{u} = \frac{\eta}{(A-\eta \sqrt{B})\sqrt{\delta}} ,
\end{align}
provided $A>\eta \sqrt{B}$. 
\end{thm}
%

The representations $\dot\vecx$ obtained from the problem $(\text{P}^{\,\varepsilon}_\infty)$ are guaranteed to satisfy~\fref{eq:maincondition}, as the proof exploits properties of its solution.
In addition, \fref{thm:main_result} shows that if a frame $\dict$ satisfies (i) $A>\eta \sqrt{B}$ and (ii) $\delta>0$, then one can compute democratic representations for every signal vector~$\vecy$ by solving $(\text{P}^{\,\varepsilon}_\infty)$. 
In addition, the condition $A>\eta \sqrt{B}$ indicates that the use of Parseval frames is beneficial in practice, i.e., leads to democratic representations with smaller $\ellinf$-norm---an observation that was made empirically by Fuchs \cite{Fuchs11b}; corresponding simulation results are provided in \fref{sec:simulation}.
In order to achieve representations having provably small $\ellinf$-norm (close to $1$), one is typically interested in finding frames satisfying the UP with small~$\eta$ and large~$\delta$.
Both properties can be achieved simultaneously for certain classes of frames (see \cite{LV10} and \fref{sec:UPsatisfyingmatrices} for corresponding examples).

We note that \fref{thm:main_result} improves upon the results in~\cite{LV10}. In particular, the bound $\widetilde{K}_\text{u}$ in \fref{eq:maincondition} is strictly smaller than the bound obtained in \cite[Thms.\ 3.5 and~3.9]{LV10}. 
To see this, consider the case of \dict being a Parseval frame and $\varepsilon=0$; this enables us to establish the following relation between the upper democracy bound $\widetilde{K}_\text{u}$ in \fref{eq:maincondition} and the bound $K$ from \cite[Thm.~3.5]{LV10}:
\begin{align*} 
\widetilde{K}_\text{u} = \frac{\eta}{(1-\eta )\sqrt{\delta}} < \frac{1}{(1-\eta )\sqrt{\delta}}=K.
\end{align*}
The strict inequality follows from the fact that $\eta$ is required to be smaller than one, which is a consequence of $A>\eta \sqrt{B}$. Hence, by solving $(\text{P}^{\,\varepsilon}_\infty)$ rather than using the algorithm proposed in~\cite{LV10}, we arrive at an upper bound that is more tight (i.e., by a factor of $\eta$). 
For approximate Parseval frames satisfying $A=1-\xi$ and $B=1+\xi$ with $0\leq\xi<1$, the upper democracy bound in \fref{eq:maincondition} continues to be superior to that in \cite[Thm.~3.9]{LV10}.
Furthermore, \fref{thm:main_result} also encompasses approximate representations ($\varepsilon>0$) and the case of complex-valued vectors and frames, which is in contrast to the results developed in \cite{LV10}.

%
%

\subsection{PAPR Properties of Democratic Representations}
\label{sec:PAR}

\sloppy

\subsubsection{PAPR reduction via democratic representations}

Democratic representations can be used to (often substantially) reduce a signal's dynamic range, which is typically characterized in terms of its PAPR (or ``crest factor'') defined below. 
For example, the transmission of information-bearing signals over frequency-selective channels typically requires sophisticated equalization schemes at the receive side.
Orthogonal frequency-division multiplexing (OFDM)~\cite{NP00} is a well-established way of reducing the computational complexity of equalization (compared to conventional  equalization schemes).
Unfortunately, OFDM signals are known to suffer from a high PAPR, which requires linear RF components (e.g.,  power amplifiers).
Since linear RF components are, in general, more costly and less power efficient compared to their non-linear counterparts, practical transceiver implementations often deploy sophisticated PAPR-reduction schemes~\cite{HL05}. 
Prominent approaches for reducing the PAPR  exploit either certain reserved OFDM tones~\cite{IS09} or the excess degrees-of-freedom in large-scale multi-antenna wireless systems~\cite{SL12}. As we will  show next, democratic representations computed via $(\text{P}^{\,\varepsilon}_\infty)$ have intrinsically low PAPR.

\fussy

We start by defining the PAPR of arbitrary vectors $\vecx$. 

%
\begin{defi}[Peak-to-average power ratio]
Let \mbox{$\vecx\in\complexset^N$} be a nonzero vector. Then, the peak-to-average power ratio~(PAPR) (or ``crest factor'') of $\vecx$ is defined as 
\begin{align} \label{eq:pardefinition}
 \textit{PAPR}(\vecx) = \frac{N\norminf{\vecx}^2}{\normtwo{\vecx}^2}.
\end{align}
\end{defi}

Note that for arbitrary vectors $\vecx\in\complexset^N$,  the PAPR satisfies the following inequalities:
\begin{align} \label{eq:simpleparbound}
1 \leq \textit{PAPR}(\vecx) \leq N,
\end{align} 
an immediate consequence of standard norm bounds. The lower bound (best case) is achieved for signals having constant amplitude (or modulus), whereas the upper bound (worst case) is achieved by vectors having a single nonzero entry.
As we will show next, the worst-case PAPR of signal representations obtained through~$(\text{P}^\varepsilon_\infty)$ is typically much smaller than the upper bound in \fref{eq:simpleparbound} suggests.
To show this, we  next bound the PAPR of signal representations obtained through $(\text{P}^\varepsilon_\infty)$ with the aid of (i)  the democratic $\ell_\infty$/$\ell_2$-norm inequality in \fref{eq:fancypantsnormbound} or (ii) the upper democracy bound in~\fref{eq:maincondition}.



The following PAPR bound only depends on the dimensions of the full-spark frame~$\bD$. 
\begin{thm}[Full-spark PAPR bound] \label{thm:parbound1}
Let $\dict\in\complexset^{M\times N}$ be a full-spark frame. Then, the PAPR of every signal representation $\dot\vecx$ obtained by solving~$(\text{P}^{\,\varepsilon}_\infty)$ satisfies 
\begin{align} \label{eq:parbound1}
 \textit{PAPR}(\dot\vecx) \leq \frac{N}{N-M+1}
\end{align}
for vectors $\vecy\in\complexset^M$ satisfying $\normtwo{\vecy} \neq 0$ and $\varepsilon<\normtwo{\vecy}$.
\end{thm}
\begin{IEEEproof}
The proof directly follows from \fref{lem:alwaysDem} and the PAPR definition~\fref{eq:pardefinition} by replacing~$\normtwo{\dot\vecx}^2$ by $(N-M+1)\norminf{\dot\vecx}^2$. 
\end{IEEEproof}

\fref{thm:parbound1} implies that $\ellinf$-norm minimization can be used as a practical and efficient substitute for minimizing the PAPR in \eqref{eq:pardefinition} \emph{directly}, which is difficult to achieve in practice.  
In addition, we observe that frames with large redundancy parameter $\lambda=N/M$ enable the computation of representations with arbitrary low PAPR (by increasing the dimension $N$). To see this, let $\lambda\to\infty$, which results in the following asymptotic bound:
\begin{align*} 
\textit{PAPR}(\dot\vecx) \leq ({1+1/N})^{-1}.
\end{align*}

The following theorem provides a PAPR bound for frames that satisfy the UP with parameters $\eta$, $\delta$  (see \fref{app:parbound} for the proof). 
\begin{thm}[UP-based PAPR bound] \label{thm:parbound}
Let $\dict\in\complexset^{M\times N}$ be a frame with the upper democracy bound~$\widetilde{K}_u$ in \fref{eq:maincondition}. Then, the PAPR of every signal representation $\dot\vecx$ obtained via~$(\text{P}^{\,\varepsilon}_\infty)$ satisfies the following bound:
\begin{align} \label{eq:parbound}
 \textit{PAPR}(\dot\vecx) \leq \widetilde{K}_\text{u}^2B
\end{align}
for vectors $\vecy\in\complexset^M$ with $\normtwo{\vecy} \neq 0$ and $\varepsilon<\normtwo{\vecy}$.
\end{thm}

\fref{thm:parbound} reveals that frames satisfying the UP and having a small upper democracy bound $\widetilde{K}_\text{u}$ are particularly effective in terms of reducing the PAPR. 
A practically relevant example of frames satisfying these properties are randomly subsampled discrete Fourier transform (DFT) matrices, which naturally appear in OFDM-based tone-reservation schemes for PAPR reduction (see, e.g., \cite{IS09} for the details). A corresponding application example is shown below in \fref{sec:applicationexample}.
%

%

It is worth mentioning that the PAPR bounds \eqref{eq:parbound1} and \eqref{eq:parbound} do \emph{not} depend on the approximation parameter~$\varepsilon$. Hence, in practice, an increase in $\varepsilon$ (as long as $\varepsilon<\normtwo{\vecy}$) is expected\footnote{We note that our own experiments show virtually no impact of the approximation parameter $\varepsilon$ to the PAPR, which confirms this behavior empirically.} to not affect the PAPR of the democratic representation $\dot\vecx$, which is in contrast to  \eqref{eq:kashinbounds}.

\subsubsection{Transmit power increase}

If democratic representations are used for PAPR reduction, e.g., in an OFDM-based communication system, then it is important to realize that transmitting~$\dot\vecx$ instead of the minimum-power (or least squares) solution $\hat\vecx$ obtained from 
\begin{align*}
(\text{P}^{\,\varepsilon}_2) \quad \underset{\tilde\vecx}{\text{minimize}}\,\, \normtwo{\tilde\vecx} \quad \text{subject to}\,\, \normtwo{\vecy-\dict\tilde\vecx} \leq \varepsilon,
\end{align*}
may result in a larger transmit power.
Therefore, it is of practical interest to study the associated power increase (PI), defined as
\begin{align} \label{eq:powerincreasedef}
\textit{PI}= \frac{\normtwo{\dot\vecx}^2}{\normtwo{\hat\vecx}^2},
\end{align}
when transmitting democratic representations $\dot\vecx$ instead of the least-squares representation $\hat\vecx$ of $(\text{P}^{\,\varepsilon}_2)$. The following result provides an upper bound on the PI (see \fref{app:gainbound} for the proof).
\begin{thm}[Power increase] \label{thm:gainbound}
Let $\dict\in\complexset^{M\times N}$ be a frame with the upper frame bound $B$ and upper democracy constant $K_u$. 
Then, the power increase, $\textit{PI}$, in \fref{eq:powerincreasedef} of every signal representation $\dot\vecx$ obtained by solving~$(\text{P}^{\,\varepsilon}_\infty)$ satisfies
\begin{align} \label{eq:powerincrease}
1\leq \textit{PI} \leq \widetilde{K}_\text{u}^2B
\end{align}
for vectors $\vecy\in\complexset^M$ satisfying $\normtwo{\vecy} \neq 0$ and $\varepsilon<\normtwo{\vecy}$.
\end{thm}

It is interesting to observe that the RHS of the bound \fref{eq:powerincrease} in \fref{thm:gainbound} coincides with the RHS of the PAPR bound~\fref{eq:parbound}. 
As a consequence, the use of frames that yield good PAPR reduction properties also guarantee a small power increase compared to directly transmitting a least-squares representation. 

\subsection{The $\ell_\inftytilde$-Norm and Its Implications}
\label{sec:devotedtodominikseethaler}

In certain applications, one might be interested in minimizing the $\ell_{\widetilde\infty}$-norm rather than the $\ell_\infty$-norm of the signal representation. 
Such representations can be useful if the PAPR of both the real and imaginary parts need to be minimized individually (see, e.g., \cite{SL12}).
In order to derive properties of signal representations obtained by solving  
\begin{align*}
(\text{P}^{\,\varepsilon}_\inftytilde) \quad \underset{\tilde\vecx\in\complexset^N}{\text{minimize}}\,\, \norminftilde{\tilde\vecx} \quad \text{subject to}\,\, \normtwo{\vecy-\dict\tilde\vecx} \leq \varepsilon,
\end{align*}
we can use the following inequalities developed in~\cite[Eq.~78]{Seethaler10}:
\begin{align} \label{eq:normrelation}
\frac{1}{\sqrt{2}}\norminf{\vecx} \leq  \norminftilde{\vecx}\leq \norminf{\vecx}.
\end{align}
These inequalities imply that all properties derived from the original problem $(\text{P}^{\,\varepsilon}_\infty)$ hold as well for democratic representations obtained by $(\text{P}^{\,\varepsilon}_\inftytilde)$ up to a factor of at most two. 
%

%% file: 4frames.tex

\section{Frames that Enable Democratic Representations}
\label{sec:UPsatisfyingmatrices}

As shown in \cite{LV10}, random orthogonal matrices, random partial DFT matrices, and random sub-Gaussian matrices satisfy the UP in \fref{def:upforframes} with high probability. Hence, matrices drawn from such classes are particularly suitable for the computation of {democratic representations} with small $\ell_\infty$-norm and for applications requiring low PAPR.
As an example, we briefly restate a result obtained in \cite{LV10} for matrices whose entries are  chosen i.i.d.\  sub-Gaussian. 

\sloppy 

\begin{defi}[Sub-Gaussian RV {\cite[Def.~4.5]{LV10}}]
A random variable $X$ is called sub-Gaussian with parameter $\beta$ if 
\begin{align*}
\Pr\{\abs{X}>u\} \leq \exp\!\left(1-{u^2}/{\beta^2}\right) \quad \text{for all} \quad u>0.
\end{align*}
\end{defi}

\fussy

For matrices having i.i.d.\ sub-Gaussian entries, the following result has been established in \cite{LV10}. 
\begin{thm}[{\!\!\cite[{\!Thm.\,4.6}]{LV10}}:\,UP for sub-Gaussian Matrices] \label{thm:LV10resultmatrix}
Let \dicta be a $M\times N$ matrix whose entries are i.i.d.\ zero-mean sub-Gaussian RVs with parameter $\beta$. Assume that $\lambda=N/M$ for some $\lambda\geq2$. Then, with probability at least $1-\lambda^{-M}$, the random matrix $\dict=\frac{1}{\sqrt{N}}\dicta$ satisfies the UP with parameters
\begin{align*}
\eta = C_0 \beta \sqrt{\frac{\log(\lambda)}{\lambda}}\quad \text{and} \quad \delta=\frac{C_1}{\lambda},
\end{align*}
where $C_0$, $C_1>0$ are absolute constants. 
\end{thm}

\fref{thm:LV10resultmatrix} implies that, for random sub-Gaussian matrices, the UP with parameters $\eta$ and $\delta$ is  satisfied with high probability. Moreover, the UP parameters $\eta$, $\delta$ only depend on the redundancy $\lambda=N/M$ of \dict.
Since $\dict=\frac{1}{\sqrt{N}}\dicta$ is not, in general, a tight frame, it was furthermore shown in \cite[Cor.~4.9]{LV10} that \dict is a so-called approximate Parseval frame with high probability, i.e., \dict satisfies the frame bounds $A=1-\xi$ and $B=1+\xi$ for some small $\xi>0$.
Hence, random sub-Gaussian matrices can be used to efficiently compute democratic representations with democracy bounds $K_\text{l}$ and $K_\text{u}$ in~\fref{eq:lowerkashinbound} and \fref{eq:maincondition} by solving $(\text{P}^{\,\varepsilon}_\infty)$. 

Reference \cite{LV10} established results similar to that of \fref{thm:LV10resultmatrix} for random orthogonal and random partial DFT matrices. Partial (or randomly sub-sampled) DFT matrices have two key advantages (over sub-Gaussian matrices): (i) they are Parseval frames, which typically yield better democracy bounds  (see~\fref{eq:maincondition} and \fref{sec:simulation} for numerical experiments), and (ii) the product of a vector with the matrix~$\bD$ or its Hermitian transpose~$\bD^H$ can be computed at low computational complexity, i.e., with  roughly $N\log_2(N)$ operations using fast Fourier transforms. The latter property is of significant practical relevance as it enables one to compute democratic representations with low computational complexity; the next section details new algorithms for solving $(\text{P}^\varepsilon_\infty$) that are able to exploit such fast transforms.

%% file: 5algorithm.tex
\section{Efficient Algorithms for Solving $(\text{P}^\varepsilon_\infty$)}
\label{sec:algorithmyipee}

In order to compute the solution to $(\text{P}^{\,\varepsilon}_\infty)$, general-purpose solvers for convex optimization problems can be used (see, e.g., \cite{BV04,cvx}).
For large-dimensional problems, however,  more efficient methods become necessary. 
A Lagrange formulation of $(\text{P}^{\,\varepsilon}_\infty)$ that leads to a computationally more efficient method, called the fast iterative truncation algorithm (FITRA), was proposed in~\cite{SL12}. However, to solve~$(\text{P}^{\,\varepsilon}_\infty)$ exactly, new algorithms are required. 

Efficient methods for solving \primal{} should be capable of exploiting fast transforms for computing $\dict\vecp$ and $\dict^H\vecq$ (for two vectors $\vecp$ and $\vecq$ of appropriate dimension).
Hence, we next propose two new algorithms that directly solve  \primal{} and are able to exploit fast transforms. The first method, referred to as CRAM (short for \underline{c}onvex \underline{r}eduction of \underline{am}plitudes) directly solves \primal{} at low computational complexity. The second method, referred to as CRAMP (short for CRAM for Parseval frames) is particularly suited for Parseval frames and for~$\varepsilon=0$, which results in even lower computational complexity than CRAM. 

\subsection{CRAM: Convex Reduction of Amplitudes}

To solve \primal{}, we use the adaptive primal-dual hybrid gradient (PDHG) scheme proposed in~\cite{GEB13}. To this end, we rewrite \primal{} as the following constrained convex program:
\begin{align*}
(\text{P}^{\,\varepsilon}_\infty) \quad  \left\{
  \begin{array}{ll}
  \underset{\tilde\vecx\in\complexset^N,\vecv\in\complexset^M}{\text{minimize}}& \|\tilde\vecx\|_\infty  \\[0.2cm]
   \text{subject to}& \vecv = \vecy-\dict\tilde\vecx, \,\, \|\vecv\|_2\le \varepsilon.
   \end{array}
   \right.
\end{align*}
The constraint $\|\vecv\|_2\le \varepsilon$ can be removed by introducing the characteristic function $\chi_\varepsilon(\vecv),$ which is zero when \mbox{$\|\vecv\|_2\le\varepsilon$} and infinity otherwise.  Additionally, we enforce the linear constraint $\vecv = \vecy-\dict \tilde\vecx$ using the Lagrange multiplier vector~$\boldsymbol\lambda\in\complexset^M$, which  yields the (equivalent) saddle-point formulation
    \begin{align*}
  \underset{\boldsymbol\lambda\in\complexset^M}{\text{max}}\,\, \underset{\tilde\vecx\in\complexset^N,\vecv\in\complexset^M}{\text{min}} \quad \|\tilde\vecx\|_\infty + \langle \dict\tilde\vecx-\vecv-\vecy, 
  {\boldsymbol\lambda} \rangle + \chi_\epsilon(\vecv),
\end{align*}
where $\langle\cdot,\cdot\rangle$ denotes the inner product.  
We emphasize that a saddle point of this problem formulation corresponds to a minimizer of \primal{}. 

We compute the saddle point of this formulation using the PDHG scheme detailed in \fref{alg:cram}. Note that the operators $\max\{\cdot,\cdot\}$ and $\text{abs}(\cdot)$, as well as the division operation $./$ on line 4 of \fref{alg:cram} operate element-wise on vector entries. 
\fref{alg:cram} converges for constant step-sizes $\tau,\sigma\in(0,\infty)$ satisfying $\tau\sigma (\|\dict\|_{2,2}^2+1)<1$ (see \cite{GEB13} for the details). To achieve fast convergence of CRAM in \fref{alg:cram}, we adaptively select the step-size parameters $\tau$, $\sigma$ using the recently developed method proposed in \cite{GEB13}.
We conclude by noting that CRAM is advantageous over other splitting methods, such as ADMM \cite{GOSB12}, which require the solution of computationally complex minimization sub-steps, such as  the solution of (possibly) high-dimensional least-squares problems.

\begin{algorithm}[t] 
\caption{CRAM: Convex Reduction of Amplitudes}\label{alg:cram}
\begin{algorithmic}[1]
\State\textbf{inputs}: {$\vecx_0\in \complexset^N,\vecy,\vecv_0,\boldsymbol{\lambda}_0  \in \complexset^M$, $\tau, \sigma \in (0,\infty)$, $k=0$} 
\While{$\text{not converged}$}
\State $\vecx_{k+1} \gets \textsc{ProxInf}( \vecx_k-\tau \dict^T\boldsymbol{\lambda}_k  ,\tau)$
\State $\vecv_{k+1} \gets \varepsilon ({\vecv_k-\tau\boldsymbol{\lambda}_k})./{\mathrm{max}\{\mathrm{abs}(\vecv_k-\tau\boldsymbol{\lambda}_k),\varepsilon \}} $
\State $\boldsymbol{\lambda}_{k+1} \gets \boldsymbol{\lambda}_k+ \sigma (\dict\vecx_{k+1}-\vecv_{k+1}-\vecy)$
\State $k\gets k + 1$
\EndWhile
\end{algorithmic}
\end{algorithm}

CRAM, as detailed in \fref{alg:cram}, requires the evaluation of the \emph{proximal} operator of the $\ell_\infty$-norm, which is given by
\begin{align} \label{eq:InfProx}
  \textsc{ProxInf}(\vecz,\tau) = \underset{\tilde\vecx\in\complexset^N}{\text{arg\,min}} \, \|\tilde\vecx\|_\infty +\frac{1}{2\tau}\|\tilde\vecx-\vecz \|^2_2,
\end{align}
where $\tau>0$. 
The minimization in  \eqref{eq:InfProx} does not have a closed form solution.  Nevertheless, one can exactly compute $\textsc{ProxInf}(\vecz,\tau)$ with low computational cost using the program detailed in \fref{alg:prox}.   


\begin{algorithm}[t]
\caption{Proximal Operator for the $\ellinf$-Norm}\label{alg:prox}
\begin{algorithmic}[1]
\Procedure{ProxInf}{$\vecz$, $\tau$}
\State \textbf{inputs}: {$\vecz \in \complexset^N,$ $\tau \in (0,\infty)$}
\State $ \veca \gets \mathrm{abs}(\vecz)$ 
\State $ \vecs \gets \mathrm{sort}(\veca, $`descending'$)$  
\For{$k = 1,\ldots,N$}
\State $c_k \gets \frac{1}{k}\sum_{i=1}^k (s_i  - \tau)$
\EndFor
\State $\alpha \gets\max\big\{0, \max_i\{{c_i}\}\big\}$
\For{$k = 1,\ldots,N$}
\State $u_k \gets \min\{a_k,\alpha\}\sign(z_k)$
\EndFor
\State \textbf{return} $\vecu$
\EndProcedure
\end{algorithmic}
\end{algorithm}

\subsection{CRAMP: CRAM for Parseval Frames}
The CRAM algorithm detailed above is suitable for arbitrary Frames and approximation parameters~\mbox{$\varepsilon\geq0$}.
We next detail an algorithm that is computationally more efficient than CRAM for the special case of Parseval frames and $\varepsilon=0$. 

CRAMP  (short for CRAM for Parseval frames) directly solves the complex-valued version of $(\text{P}_\infty)$ by alternating between projections onto the linear constraint $\dict \vecx=\vecy$ and the evaluation of the \emph{proximal} operator of $\ell_\infty$-norm as in \eqref{eq:InfProx}.
For general frames, the projection of $\vecx$ onto the linear constraint $\dict \vecx=\vecy$ is given by
\begin{align*}
\Pi(\vecx) = \vecx-\dict^H(\dict\dict^H)^{-1}(\dict\vecx-\vecy).
\end{align*}
This projection, however, requires the computation of the inverse $(\dict\dict^H)^{-1}$, which may result in significant computational costs. 
When $\dict$ is a Parseval frame, we have \mbox{$\dict\dict^H=\bI_{M\times M}$}. Consequently, the above projection simply corresponds to
\begin{align} \label{eq:frameprojection}
   \Pi(\vecx) = \vecx-\dict^H(\dict\vecx-\vecy),
 \end{align}
 which can be carried out at low computational complexity. Hence, CRAMP is particularly suited for Parseval frames.\footnote{Note that the projection \eqref{eq:frameprojection} can easily adapted to the case of tight frames, i.e., where $A=B$. The resulting projection operator is simply given by $\Pi(\vecx) = \vecx-A^{-1}\dict^H(\dict\vecx-\vecy)$.} 

CRAMP is obtained by applying Douglas-Rachford splitting to the equivalent optimization problem
\begin{align*}
(\text{P}_\infty) \quad \underset{\tilde\vecx\in\complexset^N}{\text{minimize}}\,\,\|\tilde\vecx\|_\infty + \chi_\Pi(\tilde\vecx),
\end{align*}
where the  proximal of the indicator function  $\chi_\Pi(\tilde\vecx)$ is simply the projection
onto the constraint $\dict\tilde\vecx=\vecy$. By using the $\ellinf$-norm proximal operator~\eqref{eq:InfProx} and the projection operator for Parseval frames~\eqref{eq:frameprojection}, we arrive at CRAMP as summarized in \fref{alg:dr}.
We note that convergence of Douglas-Rachford splitting has been proved for arbitrary convex functions and any positive stepsize $\tau>0$ \cite{EB92}.
 
\begin{algorithm}[t] 
\caption{CRAMP: CRAM for Parseval Frames} \label{alg:dr}
\begin{algorithmic}[1] 
\State \textbf{inputs}: {$\vecz_0  \in \complexset^N$, $\tau \in (0,\infty)$}, $k=0$
\While{$\text{not converged}$}
\State $\hat \vecx_{k} \gets \textsc{ProxInf}(\vecz_k,\tau)$
\State $\hat \vecz_{k} \gets 2 \hat \vecx_k-\vecz_k$
\State $\vecx_{k+1} \gets \Pi(\hat\vecz_k)$
\State $\vecz_{k+1} \gets \vecz_k+ \vecx_{k+1}-\hat \vecx_{k}$
\State $k\gets k+1$
\EndWhile
\end{algorithmic}
\end{algorithm}

The CRAMP algorithm exhibits a practically relevant advantage over CRAM:  Every iterate produced by the CRAMP algorithm is \emph{feasible} (i.e., $\dict \vecx_k=\vecy$ for all $k$).  This property is particularly important in real-time signal processing systems where algorithms are terminated after a pre-determined number of iterations to meet tight throughput constraints.  Because all iterates are feasible, CRAMP is guaranteed to terminate with an exact  representation $\vecx$ of the signal vector $\vecy,$ regardless of whether convergence to a \emph{minimum} $\ellinf$-norm solution has been reached.

%% file: 6results.tex

\section{Numerical Experiments}
\label{sec:simulation}

We next provide numerical results that empirically characterize the key properties of democratic representations shown in \fref{sec:properties}. 
In particular, we simulate a lower bound on $K_\text{u}$ in~\fref{eq:maincondition} and evaluate the PAPR behavior of solutions to $(\text{P}^{\,\varepsilon}_\infty)$ for complex i.i.d.\ Gaussian and randomly subsampled discrete Fourier transform (DFT) bases. We finally show an application example of democratic representations for PAPR reduction in an OFDM-based DVB-T2 broadcast system. 

\subsection{Impact of Frame Properties on the Upper Democracy Constant}
\label{sec:simkashinbound}

\begin{figure}[tb]
\centering
\includegraphics[width=0.90\columnwidth]{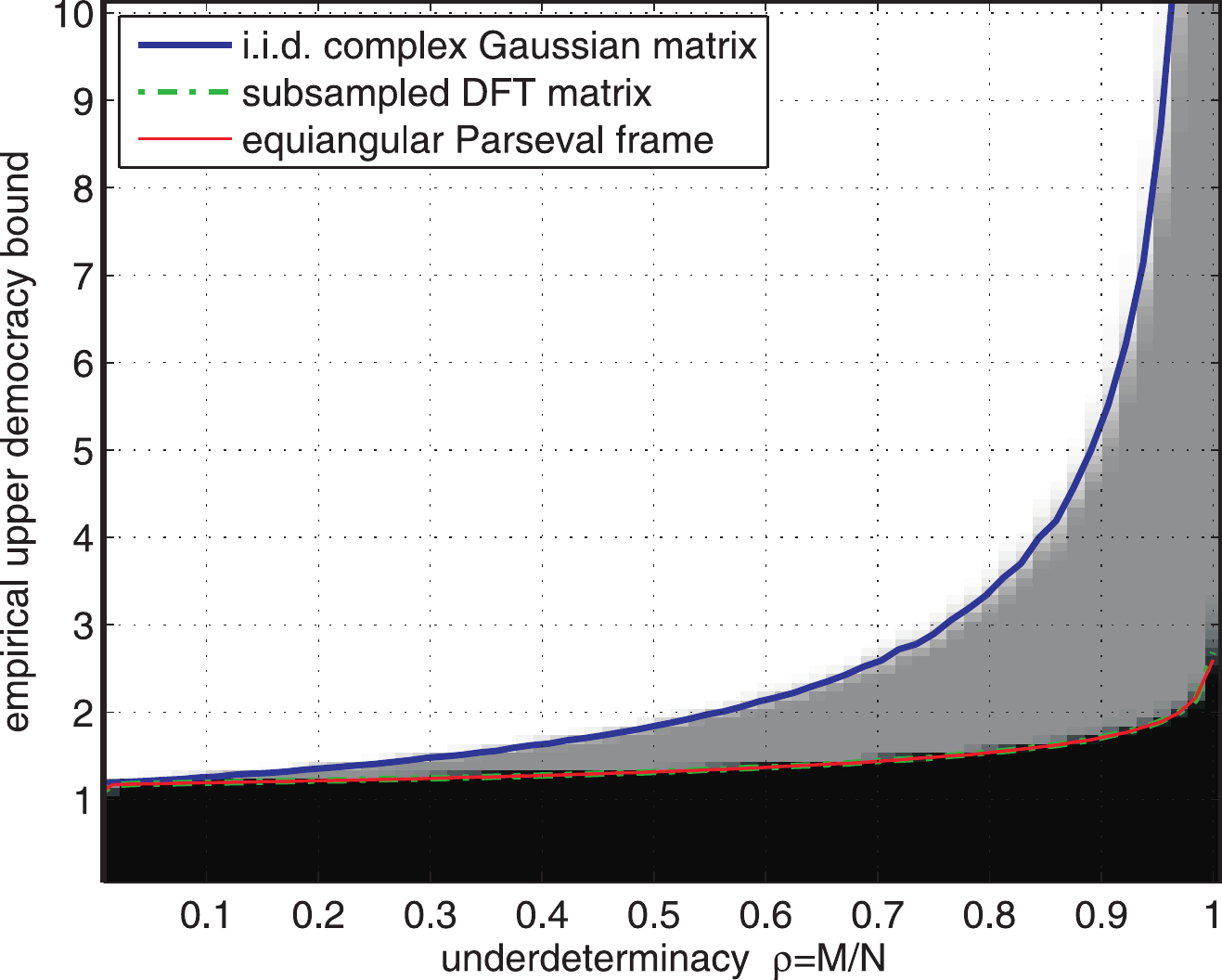}
\caption{Empirical phase diagram depending on the underdeterminancy \mbox{$\rho=M/N$} for the upper Kashin bound $K_\text{u}$ using $(\text{P}^{\,\varepsilon}_\infty)$  with $\varepsilon=0$ and for (i)  i.i.d.\ complex Gaussian matrices, (ii) randomly subsampled DFT matrices, and (iii) equiangular Parseval frames. The curves represent the individual, sharp 50\% phase-transition boundaries. (Note that the curves for subsampled DFT matrices and equiangular Parseval frames overlap.)} 
\label{fig:kashinbound}
\end{figure}

\begin{figure}[tb]
\centering
%
\includegraphics[width=0.90\columnwidth]{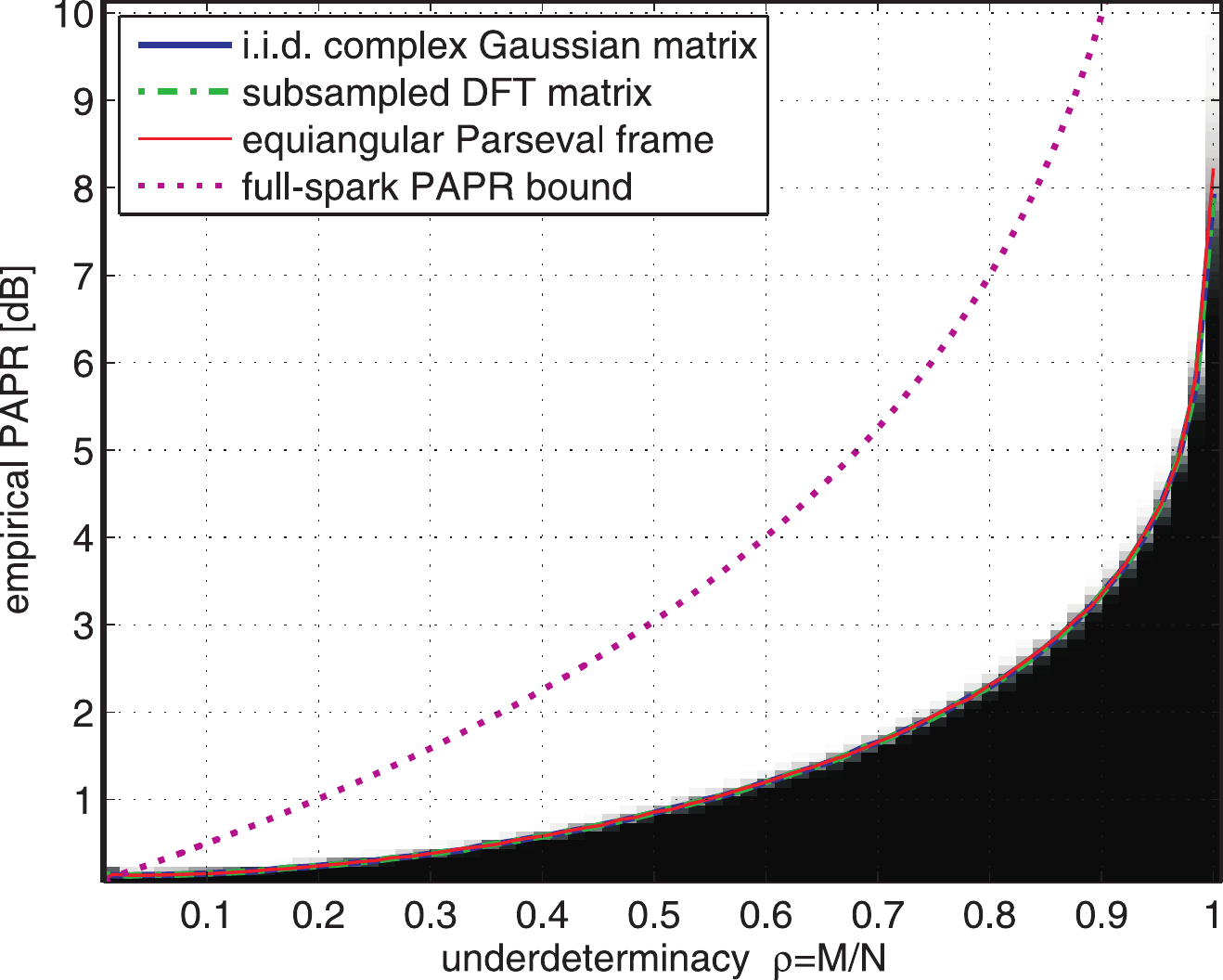}
\caption{Empirical phase diagram depending on the underdeterminancy \mbox{$\rho=M/N$} for the PAPR using $(\text{P}^{\,\varepsilon}_\infty)$  with $\varepsilon=0$ and for (i)  i.i.d.\ complex Gaussian matrices, (ii) randomly subsampled DFT matrices, and (iii) equiangular Parseval frames. The curves represent the individual, sharp 50\% phase-transition boundaries. We also show the full-spark PAPR bound~\eqref{eq:parbound1}. (Note that the curves for i.i.d.\ complex Gaussian matrices, subsampled DFT matrices, and equiangular Parseval frames overlap.)}
\label{fig:papr}
\vspace{-0.4cm}
\end{figure}

In \fref{fig:kashinbound}, we show empirical phase diagrams that characterize the upper democracy bound $K_\text{u}$ for i.i.d.\ Gaussian matrices, randomly subsampled DCT matrices, and equiangular Parseval frames constructed using the algorithm of~\cite{tropp2005}.

\subsubsection{Simulation procedure} 
\label{sec:simprocedure1}
We fix $N=512$ and vary $M$ from 1 to 512. For each measurement/dimension pair $(N,M)$, we perform 100 Monte-Carlo trials, and for each trial we generate a frame $\dict$ from each matrix/frame class specified above. We furthermore generate a complex  i.i.d.\ zero-mean Gaussian vector~$\vecy$ and normalize it to \mbox{$\normtwo{\vecy}=1$}. We use  CRAMP from \fref{sec:algorithmyipee} to compute signal representations $\dot\vecx$ from~$(\text{P}^{\,\varepsilon}_\infty)$ with $\varepsilon=0$ for each instance of $\dict$ and $\vecy$. We then compute an empirical lower bound~$\hat{K}_\text{u}$ on the upper democracy constant using the obtained representations $\dot\vecx$ for each trial as follows:
\begin{align} \label{eq:lowerboundonupperkashin}
 \hat{K}_\text{u} = \frac{\sqrt{N}\norminf{\dot\vecx}}{\normtwo{\vecy}-\varepsilon}  \leq K_\text{u} \quad \text{with} \quad \varepsilon=0.
\end{align}
We finally generate phase diagrams, which show the empirical probability for which $\hat{K}_\text{u}$ is larger or smaller than a given empirical upper democracy constant (given by the y-axis).

\subsubsection{Discussion}
The empirical phase diagram shown in \fref{fig:kashinbound} shows a sharp transition between the values of $\hat{K}_\text{u}$ that have been realized (for a given under determinacy $\rho=M/N$) and the values that were not  achieved. Moreover, we see that subsampled DFTs and equiangular Parseval frames have smaller (empirical) upper democracy constant than that of i.i.d.\ Gaussian matrices. 
This behavior is predicted by~\fref{eq:maincondition} and observed previously \cite{Fuchs11b}, and can be attributed to the fact that subsampled DFT matrices are Parseval frames, whereas i.i.d.\ Gaussian matrices are, in general, not tight frames (see also  \fref{sec:UPsatisfyingmatrices}).
Hence, the use of Parseval frames  tends to yield democratic representations with smaller $\ellinf$-norm than general (non-tight) frames, which is reflected by the upper democracy bound of \fref{eq:maincondition} that explicitly depends on the frame bounds $A$ and $B$.

\subsection{Impact of Frame Properties on PAPR}

In \fref{fig:papr}, we characterize the impact of frame properties on the PAPR of signal representations obtained by solving~$(\text{P}^{\,\varepsilon}_\infty)$. 

\subsubsection{Simulation procedure} 
We carry out a similar simulation procedure as in \fref{sec:simprocedure1}, but instead we
 compute $\textit{PAPR}(\dot\vecx)$ for each instance of $\dict$ and $\vecy$. 

\subsubsection{Discussion} 
The phase diagram shown in \fref{fig:papr} exhibits a sharp transition between the (empirical) PAPR values achieved in this simulation and the values that were not achieved. It is interesting to see that all 50\% phase transitions overlap, which is in stark contrast to the transition behavior of the upper democracy bound discussed above.
We, hence, conclude that the particular choice of the frame has a negligible impact for PAPR-reduction.
In addition, \fref{fig:papr} also shows the full-spark PAPR bound \fref{eq:parbound1}. 
We note that the PAPR bound in~\fref{thm:parbound} depends on the upper frame bound $B$, which is not reflected in this simulation; an investigation of a tighter PAPR bound is part of ongoing work. Furthermore, the gap between the 50\% phase transition and the full-spark PAPR bound appears to rather large. Nevertheless, we note that the full-spark PAPR bound does  neither depend on the signal to be represented nor on the specifics of the used frame (apart from its dimensions). Hence, one can imagine that for certain frames one might be able to construct adversarial signals whose representations exhibit high PAPR. 

\subsection{Application Example: PAPR Reduction in DVB-T2}
\label{sec:applicationexample}

We now show a simple application example of democratic representations for PAPR reduction in an OFDM-based DVB-T2 broadcast system~\cite{DVBT2}. While this example demonstrates the efficacy of democratic representations for PAPR reduction, we do not intend to provide a thorough comparison with state-of-the-art algorithms used in real-world implementations. For a more detailed discussion on this matter, we refer the interested reader to \cite{tarokh2000computation,NP00,IS09,SL12}.

\subsubsection{Algorithm details and simulation procedure} 
We consider a simplified\footnote{We ignore DVB-T2-specific OFDM frame structures, such as pilot tones. For the sake of simplicity, we generate 256-QAM symbols for all used tones.} DVB-T2 system, where we use the tones reserved for PAPR reduction to generate OFDM time-domain signals~$\vecx$ having low PAPR.
In particular, we generate the entries of the frequency-domain (signal) vector $\vecy$ by inserting i.i.d.\ random 256-QAM symbols into the data-carrying tones and by inserting $0$'s into the specified zero-tones. The set of entries in $\vecy$ containing the constellation symbols and the zero-tones is denoted by $\Omega$; the complement set $\Omega^c$ contains the tones reserved for PAPR reduction. 
 We can now write the time-domain vector~$\vecx$ as $\bF\vecx=\vecy$, where $\bF$ is a DFT matrix of appropriate size that satisfies $\bF\bF^H=\bI$. In the following experiment, we use a DFT of dimension $N=32\,768$ as specified in~\cite{DVBT2}. For this particular DFT size, we have $\abs{\Omega^c}=288$ tones reserved for PAPR reduction. 
By separating the signal vector $\vecy$ into two disjoint parts $\vecy_\Omega$ and $\vecy_{\Omega^c}$, we can rewrite the time-domain vector as $\vecx= \bF^H_\Omega\vecy_\Omega+\bF^H_{\Omega^c}\vecy_{\Omega^c}$. Hence, for the OFDM tones in $\Omega$, we have  $\vecy_{\Omega} = (\bF^H_\Omega)^H\vecx$; 
here,  $(\bF^H_{\Omega})^H$ is a subsampled DFT matrix having $M=32\,480$ rows from the set $\Omega$ and all $N=32\,768$ columns. Since~$\vecx$ is the time-domain vector to be transmitted, we can reduce its PAPR by solving the following problem:
\begin{align*}
(\text{PR}) \quad \underset{\tilde\vecx\in\complexset^N}{\text{minimize}}\,\, \norminf{\tilde\vecx} \quad \text{subject to}\,\, \vecy_{\Omega}=(\bF^H_{\Omega})^H\tilde\vecx,
\end{align*}
which is a specific instance of $(\text{P}^{\,\varepsilon}_\infty)$ with $\bD=(\bF^H_{\Omega})^H$ and $\varepsilon=0$.
By solving $(\text{PR})$ we obtain a time-domain signal $\dot\vecx$ that has  (i) low PAPR and (ii) a frequency-domain representation that corresponds to $\vecy_\Omega$ on the set of used OFDM tones. 

We note that, in practice, the time-domain signals pass through a digital-to-analog converter, which typically applies a low-pass reconstruction filter to the resulting time-domain signal. To accurately assess the PAPR of the resulting analog (filtered) time-domain signal, one typically considers the PAPR of an oversampled system, which is achieved by an appropriate zero-padding of frequency-domain vector $\vecy$ (see \cite{MNPGD10} and the references therein). 
As in \cite{MNPGD10} we compute the PAPR using $4\times$ oversampling. 

\sloppy

In the following experiments, we generate $10^5$ OFDM signals~$\vecy$ as specified above and compare the PAPR  of the following methods/algorithms: (i) conventional OFDM transmission (where $y_i=0$ for $i\in \Omega^c$); (ii) PAPR-reduced OFDM transmission using the algorithm detailed in the DVB-T2 standard\footnote{We perform $1\,000$ algorithm iterations and use a set of optimized algorithm parameters to achieve minimal PAPR.} \cite[Sec.~9.6.2.1]{DVBT2}; (iii) PAPR-reduced OFDM transmission as by solving $(\text{PR})$ via CRAMP\footnote{The maximum number of iterations is set to $1\,000$; on average, CRAMP terminates after $250$ iterations.}; (iv) PAPR-reducd OFDM transmission by solving a variant of $(\text{PR})$ via CRAMP that directly operates on the $4\times$ oversampled system. As a performance measure, we compare the complementary cumulative distribution function (CDF) of the oversampled PAPR values (in decibel) obtained in all  simulation trials~\cite{MNPGD10}.  

\fussy

\begin{figure}[tb]
\centering
\includegraphics[width=0.90\columnwidth]{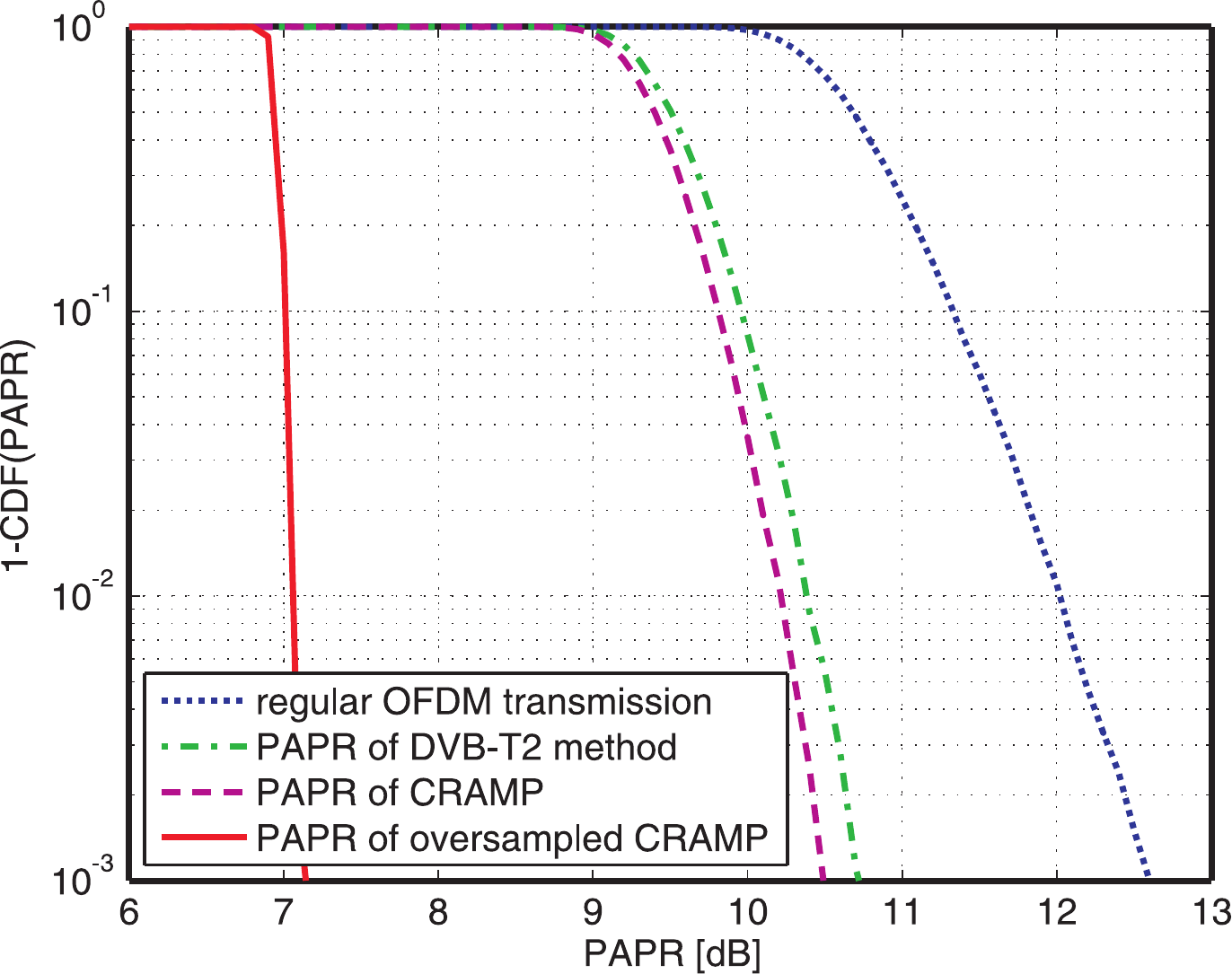}
\caption{PAPR reduction performance for a DVB-T2-based broadcast system with 32\,768 OFDM tones and 256-QAM modulation~\cite{DVBT2}. We compare the PAPR reduction performance of (i) regular OFDM transmission (no PAPR reduction use), (ii) the algorithm specified in the DVB-T2 standard, (iii) CRAMP, and (iv) CRAMP directly applied to the $4\times$ oversampled system.}
\label{fig:paprreduction}
\end{figure}

\subsubsection{Discussion} 

From \fref{fig:paprreduction}, we see that conventional OFDM transmission exhibits the largest PAPR. The algorithm in \cite{DVBT2} is able to reduce the PAPR by roughly $1.5$\,dB (corresponding to a complementary CDF of $10^{-2}$). Solving $(\text{PR})$ via CRAMP reduces the PAPR by roughly $1.7$\,dB. Solving $(\text{PR})$ directly on the oversampled system leads to a significant PAPR reduction of about $4.5$\,dB (or $3$\,dB more than conventional schemes). 
Hence, PAPR reduction using $\ellinf$-norm minimization is able to significantly outperform existing methods. 

We conclude by noting that CRAMP-based PAPR reduction exhibits, in general, higher computational complexity than the algorithm specified in the DVB-T2 standard~\cite{DVBT2}, but requires no parameter tuning. Since CRAMP does not exploit the fact that the effective matrix $(\bF^H_{\Omega})^H$ in the considered application has a very specific structure and extremely low redundancy (i.e., $\lambda\approx1.008$), we are convinced that more efficient algorithms can be developed for solving $(\text{PR})$ in this particular setting. 


%% file: appendix.tex

\section{Proof of \fref{lem:demsExist}}
\label{app:demsExist}

Suppose that $\dot \vecx$ is a nonzero solution to  \primal{} with fewer than $N-M+1$ extreme values. Without loss of generality, suppose the first $K$ entries of $\dot\vecx$ are extreme, where $K<N-M+1$. Let $\dot{\vecx}_1$ and $\dot{\vecx}_2$ be the first $K$ entries and remaining entries of  $\dot\vecx$, respectively. Similarly, let $\bD_1$ be formed by the first $K$ columns of $\bD$ and $\bD_2$ be formed by its remaining $N-K$ columns.  Note that $\bD_2$ has $M$ rows and $N-K$ columns, where $M\le N-K$. Hence, we have either (i) $\rank(\bD_2)=M$ or (ii) $\rank(\bD_2)<M\le N-K$. 

\sloppy

Case (i): $\rank(\bD_2)=M$. There exists a nonzero vector $\vecv_2$ such that $\bD_2 \vecv_2=\bD\dot{\vecx}$. Let $\dot{\vecy}_\alpha = (1-\alpha)\dot{\vecx} + \alpha[\bZero\,;\vecv_2] = [(1-\alpha)\dot{\vecx}_1\,;(1-\alpha)\dot{\vecx}_2+\alpha \vecv_2]$. We have $\bD\dot{\vecy}_\alpha = (1-\alpha)\bD\dot{\vecx}+ \alpha \bD_2 \vecv_2=(1-\alpha)\bD\dot{\vecx}+ \alpha\bD\dot{\vecx}=\bD\dot{\vecx}$, so $\dot{\vecy}_\alpha$ is a feasible solution. Since  $\|\dot{\vecx}_1\|_\infty>\|\dot{\vecx}_2\|_\infty$, there exists an $\alpha>0$ such that 
\begin{align*}\|(1-\alpha)\dot{\vecx}_1\|_\infty=\|(1-\alpha)\dot{\vecx}_2+\alpha \vecv_2\|_\infty=\|\dot{\vecy}_\alpha \|_\infty<\|\dot{\vecx}\|_\infty.
\end{align*} 
We reach a contradiction that $\dot{\vecy}_\alpha$ is a candidate solution  strictly better than~$\dot \vecx$. Therefore, this case is impossible.

\fussy

Case (ii): $\rank(\bD_2)<M\le N-K$. There exists a nonzero vector $\vecv_2$ such that $\bD_2 \vecv_2 =0$. Then,  $\dot{\vecy}_\alpha = \dot{\vecx} + \alpha[\bZero\,;\vecv_2]=[\dot{\vecx}_1\,;\dot{\vecx}_2+\alpha \vecv_2]$ satisfies $\bD\dot{\vecy} = \bD\dot{\vecx}$ for any value of $\alpha$; thus $\dot{\vecy}_\alpha $ is feasible. We can select $\alpha$ such that $\|\dot{\vecx}_2+\alpha \vecv_2\|_\infty = \|\dot{\vecx}_1\|_\infty$ since $\|\dot{\vecx}_2\|_\infty<\|\dot{\vecx}_1\|_\infty$. Then $ \|\dot{\vecy}_\alpha\|_\infty=\max\{\|\dot{\vecx}_1\|_\infty,\|\dot{\vecx}_2+\alpha \vecv_2\|_\infty\}=\|\dot{\vecx}_1\|_\infty=\|\dot{\vecx}\|_\infty.$ Therefore, $\dot{\vecy}_\alpha$ is feasible, achieves the same objective, and has at least one more extreme value than  $\dot \vecx$.

\section{Proof of \fref{lem:alwaysDem}}
\label{app:alwaysDem}

Assume for contradiction that \primal{} admits a solution $\dot\vecx$ strictly fewer than $N-M+1$ extreme values. Let $\dot{\vecx}_1$, $\dot{\vecx}_2$, $\bD_1$, and $\bD_2$ be defined the same as in \ref{app:demsExist}. Note that $\bD_2$ has $M$ rows and $N-K$ columns, where $M\le N-K$. Since $\bD$ is a full-spark frame, $\rank(\bD_2)=M$, which is  Case (i) in \ref{app:demsExist} and leads to a contradiction following the same arguments.

\section{Proof of \fref{thm:primaldualproblems}}
\label{app:primaldualproblems}

Let $\|\vecw\|_p$ and $\|\vecv\|_d$ denote the primal and dual norm of the vectors \vecw and \vecx satisfying
\begin{align*}
\|\vecw\|_p = \max_{\vecv} \left\{ \Re\big(\vecv^H\vecw\big) \colon\! \|\vecv\|_d\leq1 \right\}
\end{align*}
with $1/p+1/d=1$ and $p,d\geq1$. Then, for primal and dual norms, we have the following result \cite{BV04}: 
\begin{align} \label{eq:primaldualproblems_step0}
\min_{\vecx} \left\{\|\vecx\|_p\!-\!\Re\big(\vecz^H\dict\vecx\big)\right\} = \left\{\begin{array}{ll}
0, & \|\dict^H\vecz\|_d\leq1\\[0.2cm]
-\infty, &  \text{otherwise}.
\end{array}\right.
\end{align}

We are now ready to derive the Lagrange dual problem~$(\text{D}^{\,\varepsilon}_p)$ to the primal problem $(\text{P}^{\,\varepsilon}_p)$. To this end, we introduce the auxiliary vector $\vecr\in\complexset^M$ to rewrite $(\text{P}^{\,\varepsilon}_p)$ as
\begin{align*}
& \min_\vecx \left\{ \|\vecx\|_p \colon \normtwo{\dict\vecx-\vecy}\leq\varepsilon\right\} \\
& \quad \qquad = \min_{\vecx,\vecr}\left\{ \|\vecx\|_p \colon \dict\vecx+\vecr=\vecy, \normtwo{\vecr}\leq\varepsilon\right\}.
\end{align*}
By introducing the Lagrange dual variable $\vecz\in\complexset^M$, we obtain 
\begin{align}
 & \min_{\vecx,\vecr}\left\{ \|\vecx\|_p \colon \dict\vecx+\vecr=\vecy, \normtwo{\vecr}\leq\varepsilon\right\} \notag \\
 & = \min_{\vecx,\vecr}\max_\vecz \left\{ \|\vecx\|_p \!-\! \Re\big(\vecz^H(\dict\vecx\!+\!\vecr\!-\!\vecy) \big) \colon\!  \normtwo{\vecr}\leq\varepsilon\right\} \notag \\
  & = \max_\vecz \min_{\vecx,\vecr} \left\{ \|\vecx\|_p \!-\! \Re\big(\vecz^H(\dict\vecx\!+\!\vecr\!-\vecy) \big) \colon\!  \normtwo{\vecr}\leq\varepsilon\right\}\!. \label{eq:primaldualproblems_step1}
\end{align}
For a given $\vecz$, the inner minimization problem of \fref{eq:primaldualproblems_step1} is separable in the unknown vectors \vecx and \vecr. The optimal auxiliary vector $\vecr$ is given by 
\begin{align*}
\vecr = \left\{\begin{array}{ll}
\varepsilon \vecz/\normtwo{\vecz}, & \vecz \neq \bZero_{M\times 1} \\[0.2cm]
\bZero_{M\times 1}, & \text{otherwise},
\end{array}\right.
\end{align*} 
and, in either case, we have $\Re\big(\vecz^H\vecr\big)=\varepsilon\normtwo{\vecz}$. Together with~\fref{eq:primaldualproblems_step0}, we find that \fref{eq:primaldualproblems_step1} is equal to
\begin{align*}
  \max_\vecz \left\{ \Re\big(\vecy^H\vecz\big)-\varepsilon\normtwo{\vecz} \colon \|\dict^H\vecz\|_d\leq1\right\}\!,
\end{align*}
which corresponds to the Lagrange dual problem $(\text{D}^{\,\varepsilon}_p)$. 
Note that since in the derivation of $(\text{D}^{\,\varepsilon}_p)$ all intermediate steps hold with equality, there is no duality gap.

\section{Proof of \fref{lem:lowerkashinbound}}
\label{app:lowerkashinbound}

The proof follows from a lower bound on the value of the dual problem $(\text{D}^{\,\varepsilon}_\infty)$.
Specifically, we have 
\begin{align} \label{eq:lower_step1}
\norminf{\dot\vecx} = \max_{\vecz}\left\{\Re\!\left(\vecy^H\vecz\right) - \varepsilon\normtwo{\vecz}\colon \normone{\dict^H\vecz}\leq1\right\}\!,
\end{align}
which we bound from below by replacing the optimal solution~$\dot\vecz$ by the following feasible solution:
\begin{align} \label{eq:lower_step2}
\hat\vecz = \frac{\vecy}{\normone{\dict^H\vecy}},
\end{align}
which satisfies the constraint $\normone{\dict^H\hat\vecz}\leq1$.
Hence, inserting~\fref{eq:lower_step2} in the right-hand side (RHS) of \fref{eq:lower_step1} leads to the following lower bound:
\begin{align} \label{eq:lower_step3}
\norminf{\dot\vecx} \geq \frac{\normtwo{\vecy}^2-\varepsilon\normtwo{\vecy}}{\normone{\dict^H\vecy}}.
\end{align}
To further bound the RHS of \fref{eq:lower_step3} from below, we use standard norm bounds and the upper frame bound $B$ of \dict to compute an upper bound to $\normone{\dict^H\vecy}$ as follows:
\begin{align}\label{eq:lower_step4}
\normone{\dict^H\vecy}\leq\sqrt{N}\normtwo{\dict^H\vecy} \leq \sqrt{N B} \normtwo{\vecy}.
\end{align}
Combining \fref{eq:lower_step4} with \fref{eq:lower_step3} finally yields
\begin{align} \label{eq:lower_step5}
\norminf{\dot\vecx} \geq \frac{\normtwo{\vecy} -\varepsilon}{\sqrt{NB}}.
\end{align}

Note that in \fref{eq:lower_step2} we assumed that $\normone{\dict^H\vecy}>0$. Since $\dict$ is a frame with lower frame bound $A>0$, we have 
\begin{align*}
\normone{\dict^H\vecy}\geq \normtwo{\dict^H\vecy}\geq  \sqrt{A}\normtwo{\vecy} >0,
\end{align*}
which is satisfied whenever $\normtwo{\vecy}>0$. In the case $\normtwo{\vecy}=0$ the bound  \fref{eq:lower_step5} continues to hold.

\section{Proof of \fref{thm:main_result}}
\label{app:proof_mainresult}

The proof proceeds in two stages. First, we separate the objective function of the Lagrange dual problem $(\text{D}^{\,\varepsilon}_\infty)$ into two independent terms and second, we derive an upper bound on the $\elltwo$-norm of the solution  $\dot\vecz$ to $(\text{D}^{\,\varepsilon}_\infty)$. 

\subsection{Separating the Result of the Lagrange Dual Problem}

From the Lagrange dual problem $(\text{D}^{\,\varepsilon}_\infty)$, we have 
\begin{align} 
\norminf{\dot\vecx} &= \Re\!\left(\vecy^H\dot\vecz\right) - \varepsilon\normtwo{\dot\vecz} \notag  \leq \abs{\vecy^H\dot\vecz} - \varepsilon\normtwo{\dot\vecz}  \notag \\
& \leq \normtwo{\dot\vecz}\!\left(\normtwo{\vecy}-\varepsilon\right),\label{eq:main_step1}
\end{align}
as an immediate consequence of the Cauchy-Schwarz inequality.\footnote{Note that the bound \fref{eq:main_step1} appears to be tight for $\varepsilon=0$, i.e., we were able to construct signal and frame instances for which we have $\norminf{\dot\vecx}=\normtwo{\vecy}\normtwo{\dot\vecz}$ up to machine precision. A systematic characterization of such signal and frame instances is left for future work.}
In the remaining steps of the proof, we derive an upper bound on~$\normtwo{\dot\vecz}$ in \fref{eq:main_step1}. To this end, we first expand
\begin{align}\label{eq:main_step1a}
\normtwo{\dot\vecz} = \normtwo{(\dict\dict^H)^{-1}\dict\dict^H\dot\vecz},
\end{align}
where $\dict\dict^H$ is invertible since \dict is a frame with lower frame bound satisfying $A>0$. Application of the Rayleigh-Ritz theorem~\cite[Thm. 4.2.2]{hornjohnson} to the right-hand side (RHS) of~\eqref{eq:main_step1a} leads to the following upper bound:
\begin{align}
\normtwo{\dot\vecz}  \leq \spectralnorm{(\dict\dict^H)^{-1}} \normtwo{\dict\dict^H\dot\vecz} 
 \leq \frac{1}{A} \normtwo{\dict\dict^H\dot\vecz},  \label{eq:main_step2}
\end{align}
where the second inequality is a result of 
\begin{align*}
\spectralnorm{(\dict\dict^H)^{-1}}  = \frac{1}{\spectralnorm{(\dict\dict^H)^{-1}}}  \leq \frac{1}{A}
\end{align*}
and the assumption that $\dict$ is a frame with lower frame bound \mbox{$A>0$}.
We next derive an upper bound on $\normtwo{\dict\dict^H\vecz}$ in~\fref{eq:main_step2}.

Note that one can straightforwardly arrive at an upper bound on $\norminf{\dot\vecx}$ as follows:
\begin{align}
\normtwo{\dict\dict^H\dot\vecz} 
& \leq \spectralnorm{\dict}\normone{\dict^H\dot\vecz} \leq \spectralnorm{\dict} \label{eq:main_step2b}
\end{align}
using $\normtwo{\dict^H\dot\vecz}\leq\normone{\dict^H\dot\vecz}$ and the constraint $\normone{\dict^H\dot\vecz}\leq1$ of the dual problem $(\text{D}^{\,\varepsilon}_\infty)$. Hence, by combining \fref{eq:main_step1}, \fref{eq:main_step2}, and \fref{eq:main_step2b} one would arrive at the following result:
\begin{align}
\norminf{\dot\vecx} \leq \frac{\spectralnorm{\dict}}{A} \big(\normtwo{\vecy}-\varepsilon\big). \label{eq:main_step2b_crap}
\end{align}
This bound is, however, overly pessimistic and does not exploit additional properties of the frame \dict. Note that  for Parseval frames, the result \fref{eq:main_step2b_crap} leads to the bound $\norminf{\dot\vecx} \leq \normtwo{\vecy}-\varepsilon$.

\subsection{Refined Upper Bound}

In order to arrive at a refined bound on $\normtwo{\dict\dict^H\dot\vecz}$, we define an $N$-dimensional vector $\vecv=\dict^H\dot\vecz$ and divide its coefficients into $S=\ceil{1/\delta}$ disjoint support sets, each\footnote{Note that the last support set $\Omega_\ell$ can have a cardinality that is smaller than $\delta N$; such cases, however, leave the proof unaffected.} of cardinality $\delta N$ such that 
\begin{align*} 
\Omega_1\cup\cdots\cup\Omega_\ell \in  \{1,\ldots,N\},
\end{align*}
where $\ceil{x}$ rounds the scalar~$x\in\reals$ to the nearest integer towards infinity. 
Moreover, the magnitudes of the entries in $\vecv$ associated to set $\Omega_\ell$ are no smaller than the magnitudes associated with the sets $\Omega_k$, $k>\ell$. In other words, $\Omega_1$ contains the indices associated to the largest $\delta N$ entries in $\vecv$, $\Omega_2$ the $\delta N$ coefficients associated to the second largest entries, etc.
This partitioning scheme now allows us to rewrite $\normtwo{\dict\dict^H\dot\vecz}$ as
\begin{align*}
\normtwo{\dict\dict^H\dot\vecz} = \normtwo{\dict\sum_{i=1}^S\bP_{\Omega_i}\dict^H\dot\vecz},
\end{align*}
where the matrix $\bP_{\Omega_i}$ realizes a projection onto the set $\Omega_i$.
Application of the triangle inequality, followed by using properties of the UP with parameters $\eta$, $\delta$ leads to the following: 
\begin{align}
\normtwo{\dict\dict^H\dot\vecz} & \leq \sum_{i=1}^S\normtwo{\dict\bP_{\Omega_i}\dict^H\dot\vecz} \leq \sum_{i=1}^S\eta \normtwo{\bP_{\Omega_i}\dict^H\dot\vecz} \notag \\
 & = \eta \normtwo{\bP_{\Omega_1}\dict^H\dot\vecz} + \sum_{i=2}^S\eta \normtwo{\bP_{\Omega_i}\dict^H\dot\vecz}.
 \label{eq:main_step3}
\end{align}
Since the sets $\Omega_i$ order the entries of $\vecv=\dict^H\dot\vecz$ according to their magnitudes, we can use a  technique developed in~\cite{C08}, which states that for $i\in\{2,\ldots,S\}$ we have 
\begin{align*}
\normtwo{\bP_{\Omega_i}\vecv} \leq \sqrt{\delta N} \norminf{\bP_{\Omega_i}\vecv} \leq \frac{1}{\sqrt{\delta N}} \normone{\bP_{\Omega_{i-1}}\vecv}.
\end{align*}
This result in combination with the RHS of \eqref{eq:main_step3} leads to
\begin{align}
\normtwo{\dict\dict^H\dot\vecz} & \leq \eta \normtwo{\bP_{\Omega_1}\dict^H\dot\vecz} + \sum_{i=1}^S\frac{\eta}{\sqrt{\delta N}} \normone{\bP_{\Omega_i}\dict^H\dot\vecz} \notag \\
& = \eta \normtwo{\bP_{\Omega_1}\dict^H\dot\vecz} + \frac{\eta}{\sqrt{\delta N}} \normone{\dict^H\dot\vecz} \notag\\
& = \eta \normtwo{\bP_{\Omega_1}\dict^H\dot\vecz} + \frac{\eta}{\sqrt{\delta N}},
\label{eq:main_step4}
\end{align}
where the first equality follows from the fact that $\normone{\dict^H\dot\vecz}\leq1$ for any solution $\dot\vecz$ to the dual problem~$(\text{D}^\varepsilon_\infty)$.

We can now bound the first RHS term in \fref{eq:main_step4} as
\begin{align} 
\normtwo{\bP_{\Omega_1}\dict^H\dot\vecz} \leq \normtwo{\dict^H\dot\vecz} \leq \sqrt{B}\normtwo{\dot\vecz}
\label{eq:main_step5}
\end{align}
using the facts that (i) $\bP_{\Omega_1}$ is a projector and (ii) $\dict$ is a frame with (upper) frame bound~$B$.
By combining \fref{eq:main_step2}, \fref{eq:main_step4}, and~\fref{eq:main_step5} we arrive at
\begin{align*}
\normtwo{\dot\vecz} & \leq \frac{1}{A}  \!\left( \eta \sqrt{B}\normtwo{\dot\vecz} + \frac{\eta}{\sqrt{\delta N}}  \right),
\end{align*}
which can be rewritten as
\begin{align} \label{eq:main_step5b}
\normtwo{\dot\vecz} & \leq \frac{ \eta }{(A-\eta \sqrt{B})\sqrt{\delta N}}
\end{align}
provided that $A>\eta \sqrt{B}$ holds.
Combining \fref{eq:main_step1} with \fref{eq:main_step5b} finally yields
\begin{align} \label{eq:main_step6}
\norminf{\dot\vecx} & \leq \frac{\eta}{(A-\eta \sqrt{B})\sqrt{\delta N}} \big(\normtwo{\vecy}-\varepsilon\big),
\end{align}
which concludes the proof.
We finally note that \fref{eq:main_step6} is able to scale in $1/\sqrt{N}\normtwo{\vecy}$ for certain frames (see \fref{sec:UPsatisfyingmatrices}).

\section{Proof of \fref{thm:parbound}}
\label{app:parbound}

The proof follows from separately bounding the numerator and denominator of the definition \fref{eq:pardefinition}. 
We first bound $N\norminf{\dot\vecx}^2$ using \fref{eq:maincondition} to arrive at
\begin{align} \label{eq:parbound_step0}
 N\norminf{\dot\vecx}^2 \leq \widetilde{K}_\text{u}^2(\normtwo{\vecy}-\varepsilon)^2.
\end{align}

The second part of the proof bounds $\normtwo{\dot\vecx}^2$ from below. 
To this end, it is important to realize that 
\begin{align} \label{eq:parbound_step1}
 \normtwo{\dot\vecx} \geq \min_{\vecx} \left\{ \normtwo{\vecx} \colon\! \normtwo{\vecy-\dict\vecx}\leq\varepsilon\right\}
\end{align}
because $\dot\vecx$ satisfies $\normtwo{\vecy-\dict\dot\vecx}\leq\varepsilon$ and the RHS is the minimizer for all vectors $\vecx\in\complexset^N$ satisfying $\normtwo{\vecy-\dict\vecx}\leq\varepsilon$.
We next compute a lower bound on the RHS of \fref{eq:parbound_step1}. 
From \fref{thm:primaldualproblems} with $p=2$ and $q=2$, we have 
\begin{align} \label{eq:parbound_step1a}
&\min_{\vecx} \left\{ \normtwo{\vecx} \colon\!  \normtwo{\vecy-\dict\vecx}\leq\varepsilon\right\} \notag \\
& \qquad = \max_{\vecz} \left\{ \Re\big(\vecy^H\vecz\big) - \varepsilon\normtwo{\vecz} \colon\!\normtwo{\dict^H\vecz}\leq1\right\}.
\end{align}
Using a similar strategy as in \fref{app:lowerkashinbound}, we replace the optimal solution $\dot\vecz$ of the dual problem in \fref{eq:parbound_step1a} by the estimate
\begin{align} \label{eq:parbound_step2}
\dot\vecz = \frac{\vecy}{\normtwo{\dict^H\vecy}},
\end{align}
which satisfies the constraint $\normtwo{\dict^H\vecy}\leq 1$. Hence, by inserting the estimate~\fref{eq:parbound_step2} into the RHS of \fref{eq:parbound_step1a}, we obtain the following lower bound:
\begin{align} \label{eq:parbound_step3}
&\max_{\vecz} \left\{ \Re\big(\vecy^H\vecz\big) - \varepsilon\normtwo{\vecz} \colon\!\normtwo{\dict^H\vecz}\leq1\right\} \notag \\
& \qquad \geq \frac{\normtwo{\vecy}^2-\varepsilon\normtwo{\vecy}}{\normtwo{\dict^H\vecy}}.
\end{align}
The upper frame bound
$\normtwo{\dict^H\vecy} \leq \sqrt{B}\normtwo{\vecy}$
enables us to further bound the RHS of \fref{eq:parbound_step3} from below as
\begin{align} \label{eq:parbound_step5}
\frac{\normtwo{\vecy}^2-\varepsilon\normtwo{\vecy}}{\normtwo{\dict^H\vecy}} \geq \frac{\normtwo{\vecy}-\varepsilon}{\sqrt{B}}.
\end{align}
By combining \fref{eq:parbound_step1}, \fref{eq:parbound_step1a}, \fref{eq:parbound_step3}, and \fref{eq:parbound_step5}, we finally obtain
\begin{align} \label{eq:parbound_step6}
 \normtwo{\dot\vecx}^2 \geq \frac{\big(\normtwo{\vecy}-\varepsilon\big)^2}{B}.
\end{align}

Consequently, if $\varepsilon<\normtwo{\vecy}$, then  we can bound the PAPR of the  representation $\dot\vecx$ obtained from $(P^{\,\varepsilon}_\infty)$  using~\fref{eq:parbound_step0} and~\fref{eq:parbound_step6} as 
$\textit{PAPR}(\dot\vecx) \leq \widetilde{K}_\text{u}^2 B$.
%
Note that in \fref{eq:parbound_step2} we assumed that $\dot\vecx\neq0$, i.e., we require $A>0$ and $\normtwo{\vecy} \neq 0$. 

\section{Proof of \fref{thm:gainbound}}
\label{app:gainbound}

For the lower bound in \fref{eq:powerincrease}, we simply recall the fact that the $\elltwo$-norm of the LS solution $\normtwo{\hat\vecx}$ is, by definition of the optimization problem $(\text{P}^{\,\varepsilon}_2)$, smaller than the $\elltwo$-norm of a democratic representation $\normtwo{\dot\vecx}$. Consequently, we get the trivial bound $\textit{PI}=\normtwo{\dot\vecx}^2/\normtwo{\hat\vecx}^2\geq1$. 

To arrive at the upper bound in \fref{eq:powerincrease}, we follow closely the proof in \fref{app:parbound} and individually bound the numerator and denominator of \fref{eq:powerincrease}.
First, we bound $\normtwo{\dot\vecx}^2$ from above as
\begin{align} \label{eq:gainbound_step1}
\normtwo{\dot\vecx}^2 \leq N \norminf{\dot\vecx}^2 \leq \widetilde{K}_\text{u}^2(\normtwo{\vecy}-\varepsilon)^2,
\end{align}
which is a consequence of standard norm bounds and the results of \fref{eq:kashinbounds} and~\fref{eq:maincondition}. 
We next obtain a lower bound on the denominator $\normtwo{\hat\vecx}^2$ of \fref{eq:powerincrease}. 
To this end, we carry out the steps in \fref{eq:parbound_step1a}--\fref{eq:parbound_step6} for $\normtwo{\hat\vecx}^2$ to arrive at the following lower bound:
\begin{align*}\label{eq:gainbound_step2}
\normtwo{\hat\vecx}^2 \geq \frac{\big(\normtwo{\vecy}-\varepsilon\big)^2}{B}.
\end{align*}
Finally, combining \fref{eq:gainbound_step1} with \fref{eq:gainbound_step1} yields
$\textit{PI} \leq \widetilde{K}_\text{u}^2 B$.
We conclude by noting that the steps in \fref{eq:parbound_step1a}--\fref{eq:parbound_step6} require $\normtwo{\vecy} \neq 0$,  $A>0$, and $\varepsilon<\normtwo{\vecy}$.

%% file: 15ACHAlinf.bbl
\begin{thebibliography}{10}
\expandafter\ifx\csname url\endcsname\relax
  \def\url#1{\texttt{#1}}\fi
\expandafter\ifx\csname urlprefix\endcsname\relax\def\urlprefix{URL }\fi
\expandafter\ifx\csname href\endcsname\relax
  \def\href#1#2{#2} \def\path#1{#1}\fi

\bibitem{LV10}
Y.~Lyubarskii, R.~Vershynin, Uncertainty principles and vector quantization,
  IEEE Trans. Inf. Theory 56~(7) (2010) 3491--3501.

\bibitem{Fuchs11}
J.-J. Fuchs, Spread represenations, in: Proc. 45th Asilomar Conf. on Signals,
  Systems, and Comput., Pacific Grove, CA, USA, 2011.

\bibitem{CD02}
A.~R. Calderbank, I.~Daubechies, The pros and cons of democracy, IEEE Trans.
  Inf. Theory 48~(6) (2002) 1721--1725.

\bibitem{PK05}
M.~P\"uschel, J.~Kova\v{c}evi\'{c}, Real, tight frames with maximal robustness
  to erasures, in: Proc. IEEE Data Compression Conf. (DDC), 2005, pp. 63--72.

\bibitem{Novak2010}
C.~Novak, C.~Studer, A.~Burg, G.~Matz, The effect of unreliable {LLR} storage
  on the performance of {MIMO-BICM}, in: Proc. of 44th Asilomar Conf. on
  Signals, Systems, and Comput., Pacific Grove, CA, USA, 2010, pp. 736--740.

\bibitem{NP00}
R.~van Nee, R.~Prasad, {OFDM} for wireless multimedia communications, Artech
  House Publ., 2000.

\bibitem{IS09}
J.~Illic, T.~Strohmer, {PAPR} reduction in {OFDM} using {Kashin's}
  representation, in: Proc. IEEE 10th Workshop on Sig. Proc. Advances in
  Wireless Comm. (SPAWC), Perugia, Italy, 2009, pp. 444--448.

\bibitem{SL12}
C.~Studer, E.~G. Larsson, {PAR}-aware large-scale multi-user {MIMO-OFDM}
  downlink, IEEE J. Sel. Areas Comm. 31~(2) (2013) 303--313.

\bibitem{JFF11}
H.~J\'egou, T.~Furon, J.-J. Fuchs, Anti-sparse coding for approximate nearest
  neighbor search, arXiv:1110.3767v2.

\bibitem{Cadzow71}
J.~A. Cadzow, Algorithm for the minimum-effort problem, IEEE Trans. Autom.
  Control. 16~(1) (1971) 60--63.

\bibitem{DW97}
A.~S. Deo, I.~D. Walker, Minimum effort inverse kinematics for redundant
  manipulators, IEEE Trans. Robotics and Automation 13~(5) (1997) 767--775.

\bibitem{ZWX02}
Y.~Zhang, J.~Wang, Y.~Xu, A dual neural network for bi-criteria kinematic
  control of redundant manipulators, IEEE Trans. Robotics and Automation 18~(6)
  (2002) 923--931.

\bibitem{ZYY13}
H.~Zhang, M.~Yan, W.~Yin, One conition for all: solution uniqueness and
  robustness of $\ell_1$-synthesis and $\ell_1$-analysis minimizations, CAAM
  Technical Report 13-10, Rice University (Apr. 2013).

\bibitem{tropp2004}
J.~A. Tropp, Greed is good: {A}lgorithmic results for sparse approximation,
  IEEE Trans. Inf. Theory 50~(10) (2004) 2231--2242.

\bibitem{eladbook2010}
M.~Elad, Sparse and Redundant Representations: {From} Theory to Applications in
  Signal and Image Processing, 1st Edition, Springer, 2010.

\bibitem{SB11}
C.~Studer, R.~G. Baraniuk, Stable restoration and separation of approximately
  sparse signals, Appl. Comput. Harmon. Anal.

\bibitem{donoho2006}
D.~L. Donoho, Compressed sensing, IEEE Trans. Inf. Theory 52~(4) (2006)
  1289--1306.

\bibitem{candes2006c}
E.~J. {Cand\`es}, J.~Romberg, T.~Tao, Robust uncertainty principles: Exact
  signal reconstruction from highly incomplete frequency information, IEEE
  Trans. Inf. Theory 52~(2) (2006) 489--509.

\bibitem{candes2008a}
E.~J. {Cand\`es}, M.~B. Wakin, An introdutction to compressive sampling, IEEE
  Sig. Proc. Mag. 25~(2) (2008) 21--30.

\bibitem{baraniuk2008a}
R.~G. Baraniuk, M.~Davenport, R.~A. DeVore, M.~B. Wakin, A simple proof of the
  restricted isometry property for random matrices, Constr. Approx. 28 (2008)
  253--263.

\bibitem{DT10}
D.~L. Donoho, J.~Tanner, Precise undersampling theorems, Proc. of the IEEE
  98~(6) (2010) 913--924.

\bibitem{MR11}
O.~Mangasarian, B.~Recht, Probability of unique integer solution to a system of
  linear equations, Europ. J. of Operational Research 214~(1) (2011) 27--30.

\bibitem{chandrasekaran2012convex}
V.~Chandrasekaran, B.~Recht, P.~A. Parrilo, A.~S. Willsky, The convex geometry
  of linear inverse problems, Foundations of Computational mathematics 12~(6)
  (2012) 805--849.

\bibitem{oymak2013simple}
S.~Oymak, C.~Thrampoulidis, B.~Hassibi, Simple bounds for noisy linear inverse
  problems with exact side information, arXiv preprint:1312.0641.

\bibitem{Neustadt66}
L.~W. Neustadt, Minimum effort control systems, J. Soc. Indus. and Appl. Math.
  Ser 1~(1) (1962) 16--31.

\bibitem{Kashin77}
B.~Kashin, Sections of some finite dimensional sets and classes of smooth
  functions, Izv. Acad. Nauk SSSR 41~(2) (1977) 334--351.

\bibitem{GG84}
A.~Y. Garnaev, E.~D. Gluskin, On widths of the {Euclidean} ball, Soviet Math.
  Dokl. 30~(1) (1984) 200--204.

\bibitem{candes2006d}
E.~J. {Cand\`es}, T.~Tao, Near-optimal signal recovery from random projections
  and universal encoding strategies?, IEEE Trans. Inf. Theory 52 (2006)
  5406--5425.

\bibitem{DVBT2}
{ETSI EN 302 755}, Digital video broadcasting ({DVB}); frame structure channel
  coding and modulation for a second generation digital terrestrial television
  broadcasting system ({DVB-T2}), Tech. rep., Version 1.3.1 (Apr. 2012).

\bibitem{Seethaler10}
D.~Seethaler, H.~B\"olcskei, Performance and complexity analysis of
  infinity-norm sphere-decoding, IEEE Trans. Inf. Theory 56~(3) (2010)
  1085--1105.

\bibitem{MB12}
V.~I. Morgenshtern, H.~B\"olcskei, A short course on frame theory, CRC Press,
  Chapter in Mathematical Foundations for Signal Processing, Communications,
  and Networking, 2012.

\bibitem{ACM12}
B.~Alexeev, J.~Cahill, D.~Mixon,
  \href{http://dx.doi.org/10.1007/s00041-012-9235-4}{Full spark frames},
  Journal of Fourier Analysis and Applications 18~(6) (2012) 1167--1194.
\newblock \href {http://dx.doi.org/10.1007/s00041-012-9235-4}
  {\path{doi:10.1007/s00041-012-9235-4}}.
\newline\urlprefix\url{http://dx.doi.org/10.1007/s00041-012-9235-4}

\bibitem{Pfetsch2012}
M.~E. Pfetsch, A.~M. Tillmann, The computational complexity of the restricted
  isometry property, the nullspace property, and related concepts in compressed
  sensing, arXiv:1205.2081v2\href {http://arxiv.org/abs/1205.2081v2}
  {\path{arXiv:1205.2081v2}}.

\bibitem{PT96}
M.~E. Pfetsch, A.~M. Tillmann, Chebotarev and his density theorem, Math.
  Intelligencer 18~(26) (1996) 26--37.

\bibitem{Fuchs05}
J.-J. Fuchs, Sparsity and uniqueness for some specific under-determined linear
  systems, in: Acoustics, Speech, and Signal Processing, 2005. Proceedings.
  (ICASSP '05). IEEE International Conference on, Vol.~5, 2005, pp.
  v/729--v/732 Vol. 5.
\newblock \href {http://dx.doi.org/10.1109/ICASSP.2005.1416407}
  {\path{doi:10.1109/ICASSP.2005.1416407}}.

\bibitem{Blumensath07}
T.~Blumensath, M.~E. Davies, Sampling theorems for signals from the union of
  finite-dimensional linear subspaces, IEEE Trans. Inf. Theory 55~(4) (2009)
  1872--1882.

\bibitem{Fuchs11b}
J.-J. Fuchs, Personal communication (2011).

\bibitem{HL05}
S.~H. Han, J.~H. Lee, An overview of peak-to-average power ratio reduction
  techniques for multicarrier transmission, IEEE Wireless Comm. 12~(2) (2005)
  1536--1284.

\bibitem{BV04}
S.~Boyd, L.~Vandenberghe, Convex Optimization, Cambridge Univ. Press, New York,
  NY, USA, 2004.

\bibitem{cvx}
M.~Grant, S.~Boyd, {CVX}: Matlab software for disciplined convex programming,
  version 1.21, \url{http://cvxr.com/cvx/} (Apr. 2011).

\bibitem{GEB13}
T.~Goldstein, E.~Esser, R.~Baraniuk, {Adaptive Primal-Dual Hybrid Gradient
  Methods for Saddle-Point Problems}, Available on Arxiv.org (arXiv:1305.0546).

\bibitem{GOSB12}
T.~Goldstein, B.~O'Donoghue, S.~Setzer, R.~Baraniuk, Fast alternating direction
  optimization methods, CAM Technical Report 12--35, UCLA (2012).

\bibitem{EB92}
J.~Eckstein, D.~P. Bertsekas, On the {Douglas-Rachford} splitting method and
  the proximal point algorithm for maximal monotone operators, Mathematical
  Programming 55 (1992) 293--318.

\bibitem{tropp2005}
J.~A. Tropp, I.~S. Dhillon, R.~W. {Heath Jr.}, T.~Strohmer, Designing
  structured tight frames via an alternating projection method, IEEE Trans.
  Inf. Theory 51~(1) (2005) 188--209.

\bibitem{tarokh2000computation}
V.~Tarokh, H.~Jafarkhani, On the computation and reduction of the
  peak-to-average power ratio in multicarrier communications, IEEE Transactions
  on Communications 48~(1) (2000) 37--44.

\bibitem{MNPGD10}
M.~Mrou\'e, A.~Nafkha, J.~Palicot, B.~Gavalda, N.~Dagorne, Performance and
  implementation evaluation of {TR PAPR} reduction methods for {DVB-T2},
  Hindawi Intl. J. of Digital Multimedia Broadcasting (2010,
  doi:10.1155/2010/797393) 1--10.

\bibitem{donoho2009}
D.~L. Donoho, A.~Maleki, A.~Montanari, Message-passing algorithms for
  compressed sensing, Proc. Natl. Acad. Sci. USA 106~(45) (2009) 18914--18919.

\bibitem{hornjohnson}
R.~A. Horn, C.~R. Johnson, Matrix Analysis, Cambridge Press, New York, NY,
  1985.

\bibitem{C08}
E.~J. Cand\`es, The restricted isometry property and its implications for
  compressed sensing, C. R. Acad. Sci. Paris, Ser. I 346 (2008) 589--592.

\end{thebibliography}
